\documentclass[%
 reprint,
 amsmath,amssymb,
 aps,
]{revtex4-2}
\usepackage{graphicx}
\usepackage{makecell}
\usepackage{bm}
\usepackage{dcolumn}
\usepackage{bm}
\usepackage{hyperref}
\usepackage{float}
    \usepackage{xcolor}

\newcommand{\bB}{{\bm{B}}}

\begin{document}

\preprint{APS/123-QED}

\title{Singly heavy Cascade Baryon $\Xi^0_{c}$ \& $\Xi^-_{b}$ Spectroscopy in the Relativistic Framework of Independent Quark Model}

\author{Rameshri V. Patel}
\email{rameshri.patel1712@gmail.com}
 \affiliation{P D Patel Institute of Applied Sciences, Charusat University, Anand 388421, Gujarat, India.}
\author{Manan Shah}     
\email{mnshah09@gmail.com}
\affiliation{P D Patel Institute of Applied Sciences, Charusat University, Anand 388421, Gujarat, India.}
\author{Bhoomika Pandya}
\affiliation{Departamento de Fìsica Teòrica and IFIC, Centro Mixto Universidad de Valencia-CSIC,
Parc Cientìfic UV, C/ Catedràtico José Beltràn, 2, 46980 Paterna, Spain}

\date{\today}

\begin{abstract}
   In this work, the potential parameters of the independent quark model are systematically reduced using previously determined inputs from a broad range of baryons. The reduced-parameter formulation, developed within the relativistic Dirac formalism and employing a Martin-like potential, is then applied to the spectroscopy of the singly heavy baryons $\Xi_c^0$ and $\Xi_b^-$. This enables an explicit verification of the linear relation obtained between the potential parameters. Radially and orbitally excited state masses are calculated, and the resulting Regge trajectories are used to assign spin-parity to experimentally observed states. The states $\Xi_c^0(2923)$, $\Xi_c^0(3080)$, $\Xi_c^0(2882)$, and $\Xi_c^0(2970)$ are identified as having $J^P = \tfrac{1}{2}^-$, $\tfrac{3}{2}^+$, $\tfrac{3}{2}^-$, and $\tfrac{1}{2}^+$, respectively, while the observed $\Xi_b^-(6227)$ state is assigned $J^P = \tfrac{5}{2}^+$. To investigate the electromagnetic structure of these baryons, their magnetic moments and radiative decay widths are computed. Additionally, the two-body weak decay branching ratios of $\Xi_c^0$ are evaluated and contrasted with experimental data to assess the robustness of the approach. The two-body nonleptonic decays of $\Xi_b^-$ are also analyzed, providin
   g predictions for branching ratios that may be tested in future experiments. Overall, the results demonstrate the effectiveness of the parameter-reduction procedure and support its applicability in the spectroscopy of baryons.

\end{abstract}

                              
\maketitle


\section{Introduction}
In 1989, the $\Xi_c^{0}$ baryon was first observed in $e^{+} e^{-}$ annihilations at CLEO~\cite{CLEO:1988yda}. In contrast, the $\Xi_b^{-}$ baryon was discovered much later, in 2007~\cite{CDF:2007cgg}. Consequently, a substantial amount of experimental data exists for the $\Xi_c^{0}$, including measurements of its higher excited state masses and decay properties, whereas observational data for the $\Xi_b^{-}$ remains comparatively limited. The singly heavy baryons, therefore, serve as valuable candidates for testing and refining phenomenological models.
Recent experimental advances have significantly expanded our understanding of $\Xi_c^{0}$ decay processes. Prior to 2019, only relative branching fraction (BF) measurements were available \cite{ParticleDataGroup:2018ovx}. In 2019, Belle conducted the first absolute BF measurements for $\Xi_c^{0}$ decays using a sample of $(772 \pm 11) \times 10^6$ $B\bar{B}$ pairs collected at the $\Upsilon(4S)$ resonance \cite{Belle:2018kzz}. For $\Xi_c^0$, the measured absolute BFs include $\mathcal{B}(\Xi_c^0 \to \Xi^- \pi^+) = (1.80 \pm 0.50 \pm 0.14)\%$, $\mathcal{B}(\Xi_c^0 \to pK^- K^- \pi^+) = (0.58 \pm 0.23 \pm 0.05)\%$, and $\mathcal{B}(\Xi_c^0 \to \Lambda K^- \pi^+) = (1.17 \pm 0.37 \pm 0.09)\%$ \cite{Belle:2018kzz}. 
Belle and Belle II have also conducted relative measurements of charmed-flavor-changing and cabibbo-suppressed  decays. The measured BFs for $\Xi_c^0 \to \Lambda \bar{K}^{*0}$ and $\Xi_c^0 \to \Lambda K_S^0$ are  $(3.3 \pm 0.3 \pm 0.2 \pm 1.0) \times 10^{-3}$ and $ (3.27 \pm 0.11 \pm 0.17 \pm 0.73) \times 10^{-3}$ \cite{Belle:2021zsy, Belle:2021avh} respectively. Decay modes involving $\Sigma^0$ or $\Sigma^+$ in the final state have also been observed \cite{Belle:2021zsy, Belle:2021avh}, with BFs given by $\mathcal{B}(\Xi_c^0 \to \Sigma^0 \bar{K}^{*0}) = (12.4 \pm 0.05 \pm 0.05 \pm 0.36) \times 10^{-3}$, $\mathcal{B}(\Xi_c^0 \to \Sigma^+ \bar{K}^{*-}) = (6.1 \pm 1.0 \pm 0.4 \pm 1.8) \times 10^{-3}$, $\mathcal{B}(\Xi_c^0 \to \Sigma^0 K_S^0) = (0.54 \pm 0.09 \pm 0.06) \times 10^{-3}$, and $\mathcal{B}(\Xi_c^0 \to \Sigma^+ K^-) = (1.76 \pm 0.10 \pm 0.14 \pm 0.39) \times 10^{-3}$. Notably, the BF for $\Xi_c^0 \to \Sigma^0 \bar{K}^{*0}$ significantly exceeds that for $\Xi_c^0 \to \Lambda \bar{K}^{*0}$, a result that conflicts with SU(3) flavor symmetry predictions and dynamical model calculations \cite{Geng:2020zgr, Hsiao:2019yur, Korner:1992wi, Zenczykowski:1993jm}.

The $W$-exchange-only decay $\Xi_c^0 \to \Xi^0 K^+ K^-$ has been observed by Belle \cite{Belle:2020ito} with a total BF of approximately $0.11\%$. Amplitude analysis reveals that $(48.1 \pm 4.2)\%$ of the events proceed resonantly through $\phi \to K^+ K^-$, while $(51.9 \pm 4.2)\%$ correspond to nonresonant $\Xi^0 K^+ K^-$ production. Additional decay channels such as $\Xi_c^0 \to \Xi^0 \pi^0$, $\Xi_c^0 \to \Xi^0 \eta$ and $\Xi_c^0 \to \Xi^0 \eta'$  have been measured to be $(6.9 \pm 0.3 \pm 0.5 \pm 1.3) \times 10^{-3}$,  $(1.6 \pm 0.2 \pm 0.2 \pm 0.3) \times 10^{-3}$, and  $(1.2 \pm 0.3 \pm 0.1 \pm 0.2) \times 10^{-3}$, respectively \cite{Belle:2024ikp}. Theoretical predictions based on SU(3) breaking consistent with these measurements \cite{Zhong:2022exp}.

Cabibbo-suppressed $\Xi_c^0$ decays such as $\Xi_c^0 \to \Xi^- K^+$, $\Lambda K^+ K^-$, and $\Lambda \phi$ have been first observed by Belle \cite{Belle:2013ntc} with BFs at the order of $10^{-4}$. These suppressed modes proceed through external and internal $W$-emission diagrams in addition to $W$-exchange contributions, providing crucial insights into the interplay between strong and weak interactions. Precise predictions of branching fractions for these decays would enable rigorous validation of phenomenological models.

 The Independent Quark Model (IQM) was originally formulated by A. Kobushkin \cite{Kobushkin:1976fq} and P. Ferreira \cite{LealFerreira:1977gz} to describe quark confinement through a linear potential. In this framework, each quark inside a baryon is treated as an independent Dirac particle moving under an average confining potential centered around center of mass of the hadronic system. Subsequent refinements of the model revealed that considering the potential as an equal admixture of scalar and vector components not only preserves relativistic consistency but also allows the Dirac equation to be recast into an effective Schrödinger form \cite{Barik:1982nr, Barik:1985rm, Barik:1986gw}.  
Unlike the heavy quark effective theory (HQET) or approaches relying on \(1/m_Q\) expansions, the IQM applies a fully relativistic treatment to both light and heavy quarks, providing a unified description of baryonic systems that incorporates effects beyond HQET’s limitations.  
The confinement dynamics of quarks within baryons are often modeled using a Martin-like potential, comprising an equal mix of scalar and vector parts. This potential has been successfully employed in the relativistic formulation of the IQM to study a wide range of meson systems \cite {Shah:2014caa, Vinodkumar:2014afm, Shah:2014aly, Vinodkumar:2015blb, Shah:2014yma, Shah:2016mgq, Pandya:2021ddc}.
 Given its strong predictive success and consistency with experimental results in the study of mesons, this model has been recently extended and refined for application to various categories of baryons \cite{Patel:2023wbs, Patel:2025gbw, Patel:2024wfo, Shah:2023myt, Shah:2023mzg}. 
 
 Two notable applications of the model are worth highlighting. First, the spectroscopy of the $\Xi^{0}$ baryon has been studied within the relativistic framework of the Independent Quark Model, using a Martin-like potential consisting of an equal mixture of scalar and vector components \cite{Patel:2024wfo}. The model successfully reproduces ground-state magnetic moments, branching ratios, and asymmetry parameters for radiative weak decays such as $\Xi^{0} \to \Lambda^{0} + \gamma$ and $\Xi^{0} \to \Sigma^{0} + \gamma$, while also predicting spin-parity assignments for excited states $\Xi^0(1950)$, $\Xi^0(2130)$, and $\Xi^0(2250)$ through Regge trajectory analysis.  
In the second case, the model was applied to singly heavy baryons $\Omega_c^{0}$ and $\Omega_b^{-}$\cite{Patel:2025gbw}. Using similarly fitted potential parameters, it provided a consistent description of their mass spectra, covering both radial and orbital excitations. The model proposed spin-parity assignments for the experimentally observed $\Omega_c^{0}(3000)$, $\Omega_c^{0}(3050)$, $\Omega_c^{0}(3067)$, $\Omega_c^{0}(3120)$, and $\Omega_c^{0}(3185)$ states, as well as for newly detected $\Omega_b^{-}$ excitations. 

Additionally, magnetic moments, radiative decay widths, and nonleptonic weak decay branching ratios were calculated, showing good agreement with available data and supporting the model’s predictive reliability.
Additional baryons, such as $\Xi_{cc}^{++}$ and $\Omega_{cc}^{+}$, have also been investigated within this framework \cite{Patel:2023wbs, Shah:2023myt, Shah:2023mzg}. Consequently, a broad spectrum of baryons, ranging from doubly heavy to light systems, has been systematically studied, with their potential parameters fitted accordingly. Notably, analysis of the fitted parameters revealed a consistent trend, allowing one potential parameter to be expressed as a linear function of the other, thus reducing the number of free parameters. As a next step, it is essential to validate this relation by testing the derived potential parameters against baryons occupying the intermediate mass region—specifically those with robust experimental data. In this context, the baryons $\Xi_c^0$ and $\Xi_b^-$ were selected for detailed study.

.

This paper presents a comprehensive discussion of the general and parameter reduction methodology in Sec. \ref{sec:2}, designed to be applicable to any baryonic system. The outlined approach involves solving the Dirac equations for individual quarks within a baryon, enabling the determination of spin-averaged masses and contributions arising from spin-spin, spin-orbit, and tensor interactions. Furthermore, we calculated key static and decay properties, including the magnetic moment and radiative decay width of a baryon. The calculation for the two body weak decays of $\Xi_c^0$ and non-leptonic decays of $\Xi_b^-$ is provided in the Sec. \ref{sec:3}. The detailed conclusion is presented in Sec. \ref{sec:4}.

\section{Parameter reduction and Independent quark model Methodology}\label{sec:2}
The independent quark model was initially formulated and extensively developed for two-quark systems, namely mesons \cite{Shah:2014caa, Vinodkumar:2014afm, Shah:2014aly, Vinodkumar:2015blb, Shah:2014yma, Shah:2016mgq, Pandya:2021ddc}. Building upon this framework, we have extended the model to encompass three-quark systems \cite{Patel:2023wbs, Patel:2025gbw, Patel:2024wfo, Shah:2023myt, Shah:2023mzg}. Within this approach, the dynamics of each constituent quark are described by a Dirac equation formulated in the rest frame of the hadron. The potential incorporated in the equation exhibits a Lorentz structure characterized by an equal admixture of scalar and vector components. In recent years, the independent quark model has been successfully applied to investigate various baryons, including $\Xi^0$, $\Omega_c^0$, $\Omega_b^-$, $\Omega_{cc}^{+}$, and $\Xi_{cc}^{++}$ \cite{Patel:2025gbw, Patel:2024wfo, Shah:2023myt, Shah:2023mzg}.
The first step in this formalism involves calculating spin-averaged masses for multiquark systems. For baryons, the spin-averaged mass is given by  
\begin{eqnarray}\label{sa}
    M_{SA}^{q_1q_2q_3} = E_{q_1}^D + E_{q_2}^D + E_{q_3}^D - E_{CM}.
\end{eqnarray}
Here, $E_{CM}$ accounts for the center-of-mass corrections, effectively eliminating all contributions arising from the center-of-mass motion. The quantity $E_{q_i}^D$ ($i = 1, 2, 3$) denotes the Dirac energy corresponding to each quark within the baryon system. These energies are determined by solving the Dirac equation for a quasi-independent quark in the center-of-mass frame, which takes the form  
\begin{equation}\label{dirac}
    \left[E_q^D - \boldsymbol{\hat{\alpha}} \cdot \boldsymbol{\hat{p}} - \hat{\beta} m_q - V(r)\right]\psi_q(\vec{r}) = 0,
\end{equation}
where $E_q^D$ represents the Dirac energy of a quark, $m_q$ is the current quark mass, and $\psi_q(\vec{r})$ is the four-component spinor wave function associated with the quark.
The quarks confined within a hadron are assumed to move independently under the influence of a flavor-independent central potential. This potential is expressed through a Martin-like functional form as  
\begin{equation}\label{potential}
    V(r) = \frac{(1+\gamma_0)}{2}(\Lambda r^{0.1} + V_0),
\end{equation}  
where $\Lambda$ denotes the strength parameter of the potential and $V_0$ represents its depth.
 As discussed in \cite{Greiner:2000cwh}, the solution of Eqn.~(\ref{dirac}) for the potential given in Eqn.~(\ref{potential}) can be written as  
\begin{equation}
    \psi_q(\vec{r}) = 
    \begin{pmatrix}
    i\, g(r)\, \Omega_{jlm}\!\left(\frac{\boldsymbol{r}}{r}\right)\\
    -f(r)\, \Omega_{jl'm}\!\left(\frac{\boldsymbol{r}}{r}\right)
    \end{pmatrix}.
\end{equation}
Here, the spinor spherical harmonics $\Omega_{jlm}$ are defined as in Ref.~\cite{Greiner:2000cwh}  
\begin{equation}
    \Omega_{jlm} = \sum_{m',m_s} \big(l\,\tfrac{1}{2}\,j \rvert m'\,m_s\,m\big)\, Y_{lm'}\, \chi_{\tfrac{1}{2}m_s},
\end{equation}
with parity $\hat{P}_0 \Omega_{jlm} = (-1)^l \Omega_{jlm}$. Here, $\chi_{\tfrac{1}{2}m_s}$ denotes the eigenfunctions of $\boldsymbol{\hat{S}}^2$ and $\hat{S}_3$, while $Y_{lm'}$ represents the standard spherical harmonics.
To estimate the Dirac energies, the coupled equations for $f(r)$ and $g(r)$ are rearranged to obtain forms analogous to the ordinary differential equation (ODE) satisfied by the reduced radial component of the Schrödinger wave function \cite{Barik:1982nr},  
\begin{multline}\label{r}
        \frac{d^2R^{Sch}(r)}{dr^2} + \Bigg[m_q(E^{Sch}_q - V(r)) 
        - \frac{l(l+1)}{r^2}\Bigg] R^{Sch}(r) = 0.
\end{multline}
The radial components of the Dirac spinors satisfy the second-order ODEs,  
\begin{multline}\label{g}
        \frac{d^2g(r)}{dr^2} + \Bigg[(E^D_q + m_q)\big(E^D_q - m_q - V(r)\big)
        - \frac{k(k+1)}{r^2}\Bigg] g(r) = 0,
\end{multline}
\begin{multline}\label{f}
        \frac{d^2f(r)}{dr^2} + \Bigg[(E^D_q + m_q)\big(E^D_q - m_q - V(r)\big)
        - \frac{k(k-1)}{r^2}\Bigg] f(r) = 0,
\end{multline}
where $k$ is the eigenvalue of the operator $\hat{k} = (1 + \boldsymbol{\hat{L}} \cdot \boldsymbol{\hat{\sigma}})$, defined as  
\begin{equation}
    k = 
    \begin{cases} 
        -(l+1) = -\left(j + \frac{1}{2}\right), & \text{for } j = l + \frac{1}{2}, \\[6pt]
        \hspace{0.75cm}l = +\left(j + \frac{1}{2}\right), & \text{for } j = l - \frac{1}{2}.
    \end{cases}
\end{equation}

For the Martin-like potential given in Eqn.~(\ref{potential}), these ODEs can be transformed into forms equivalent to the Schrödinger-type ODE by introducing a dimensionless variable $\rho = \frac{r}{r_0}$, leading to the relations  
\begin{equation}\label{grho}
\frac{d^2g(\rho)}{d\rho^2} + \left[\epsilon^D - \rho^{0.1} - \frac{k(k+1)}{\rho^2}\right] g(\rho) = 0,
\end{equation}
\begin{equation}
\frac{d^2f(\rho)}{d\rho^2} + \left[\epsilon^D - \rho^{0.1} - \frac{k(k-1)}{\rho^2}\right] f(\rho) = 0,
\end{equation}
\begin{equation}
\frac{d^2R^{Sch}(\rho)}{d\rho^2} + \left[\epsilon^{Sch} - \rho^{0.1} - \frac{l(l+1)}{\rho^2}\right] R^{Sch}(\rho) = 0.
\end{equation}

Here, the parameters $\epsilon^D$ and $\epsilon^{Sch}$ are defined as  
\begin{equation}
\epsilon^D = (E^D_q - m_q - V_0)(m_q + E^D_q)^{\frac{0.1}{2.1}}
             \left(\frac{2}{\Lambda}\right)^{\frac{2}{2.1}},
\end{equation}
\begin{equation}
\epsilon^{Sch} = m_q(E^{Sch}_q - V_0)\, (m_q)^{\frac{-2}{2.1}}
                 \left(\frac{1}{\Lambda}\right)^{\frac{2}{2.1}}.
\end{equation}

For the Schrödinger case, the scale parameter is given by 
$r_0 = (m_q \Lambda)^{\frac{-1}{2.1}}$, 
while in the Dirac case, it is 
$r_0 = \left[(m_q + E^D_q)\frac{\Lambda}{2}\right]^{\frac{-1}{2.1}}$ 
\cite{Barik:1982nr}. The Schrödinger equation can be solved numerically using the procedure outlined in Ref.~\cite{W Lucha : 1998}, and the Dirac energies of the individual quarks are obtained by equating $\epsilon^D$ with $\epsilon^{Sch}$.

The parameters of the potential are determined by fitting the theoretical spin-averaged mass, obtained from Eqn.~(\ref{sa}), to the corresponding experimental spin-averaged mass of the $S$ wave. The experimental spin-averaged mass is evaluated using the relation  
\begin{equation}
M_{SA} = \frac{\sum_J (2J + 1) M_{nJ}}{\sum_J (2J + 1)},
\end{equation}  
which, for the $S$ waves of baryons, simplifies to $(M_{1/2} + 2M_{3/2})/3$. Once the potential parameters are determined, the same formulation can be employed to compute the spin-averaged masses of the excited $S$ wave states as well.

\begin{table}[h]
\caption{\label{tab:table1}Potential parameters}
\renewcommand{\arraystretch}{1.5}{
\setlength{\tabcolsep}{15pt}
\begin{tabular}{c c c }
\hline 
\hline
Baryon & $\Lambda$ $(GeV^{1.1})$ & $V_0$ $(GeV)$\\ 
\hline
$\Xi_{cc}^{++}$ & $1.150$ & $-0.960$\\
$\Omega_c^0$ & $1.250$ & $-1.011$\\
$\Omega_{cc}^{++}$ & $1.252$ & $-1.009$\\
$\Omega_b^-$ & $1.450$ & $-1.110$\\
$\Xi^0$ & $1.890$ & $-1.893$\\
\hline
\hline
\end{tabular}}
\end{table}

\begin{figure*}
\includegraphics[scale=0.85]{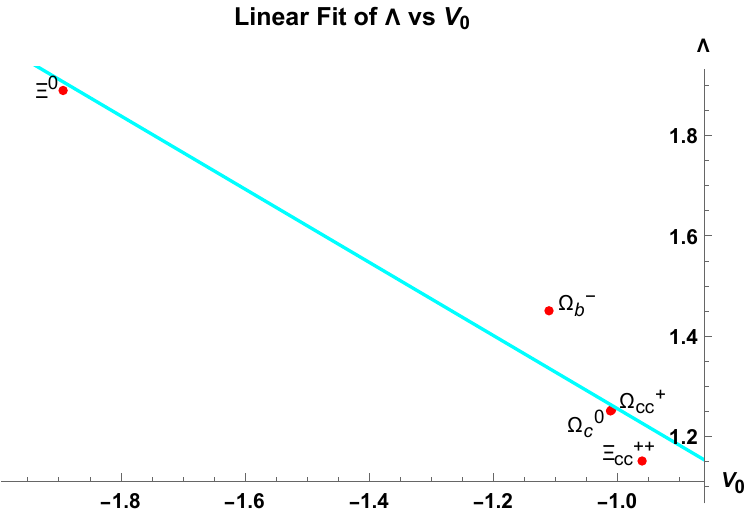}
\caption{\label{fig1} The red points represent the parameters corresponding to different baryons, while the cyan line indicates the fitted linear regression.}
\end{figure*}

The potential parameters obtained for all baryons are listed in Table~\ref{tab:table1}. A linear correlation is observed among the potential parameters, as illustrated in Fig.~\ref{fig1}, allowing one parameter to be expressed as a function of the other.  
A linear fit of $\Lambda$ as a function of $V_0$ yields the relation  
\begin{equation}\label{lambda}
\Lambda = -0.7308\,V_0 + 0.5239.
\end{equation}
This correlation effectively reduces the number of free parameters in the model, allowing only $V_0$ to be directly fitted to the experimental data for any given baryon. 

In this model, the spin-spin (hyperfine), spin-orbit, and tensor interactions are treated as perturbative corrections to the dominant central potential, following the standard formalism used in quark models. These interactions lead to fine and hyperfine splittings in the baryon spectrum (typically of the order of tens of MeV compared to the GeV-level eigenenergies) and are naturally suppressed for heavy quarks due to their inverse dependence on relativistic energy terms. This suppression justifies their perturbative treatment and subsequent inclusion as corrections to the spin-averaged mass.  

To account for spin degeneracy, the spin-spin interaction is introduced into $M_{SA}$ by considering the total spin of the three-quark system, defined as $\vec{J}_{3q} = \vec{J}_1 + \vec{J}_2 + \vec{J}_3$. The spin-spin interaction is given by  
\begin{equation}\label{vjj}
\big<V^{jj}_{q_1 q_2 q_3}(r)\big> = 
\sum_{i=1,\,i<k}^{i,k=3} 
\frac{\sigma\, 
\big<j_i.j_kJM \rvert 
\widehat{j_i}\cdot\widehat{j_k} 
\rvert j_i.j_kJM\big>}
     {(E^D_{q_i} + m_{q_i})(E^D_{q_k} + m_{q_k})},
\end{equation}  
where the total interaction is expressed as the sum of contributions from all quark pairs. Here, $\sigma$ denotes the $j$–$j$ coupling constant, whose value is determined by fitting to experimental data.
The fitted potential parameter $V_0$ and the corresponding $\Lambda$ values, obtained using the relation~(\ref{lambda}), together with other model parameters for the $\Xi_c^0$ and $\Xi_b^-$ baryons, are presented in Tables~\ref{tab:1} and~\ref{tab:2}, respectively.
 
\begin{table}[!tbh]
\begingroup  
\caption{Fitted parameters for the $\Xi_c^{0}$} \label{tab:1}
\setlength{\tabcolsep}{2pt}
\renewcommand{\arraystretch}{1.5}
\begin{tabular}{c c}
\hline
\hline
Parameter & Value (With $5\%$ variation)\\
\hline
Depth of the potential ($V_0$) & $-1.160\pm 0.058$ $GeV$  \\
Potential strength ($\Lambda$)  & $1.372 \pm 0.068$ $GeV^{1.1}$ \\
Center of mass correction  ($E_{CM}$)  & $0.302 \pm 0.015$ $GeV$ \\
$j-j$ coupling constant ($\sigma$) & $0.152 \pm 0.008$ $GeV^{3}$\\
\hline
\hline
\end{tabular}
\endgroup
\end{table}

\begin{table}[!tbh]
\begingroup
\caption{Fitted parameters for the $\Xi_b^{-}$} \label{tab:2}
\setlength{\tabcolsep}{2pt}
\renewcommand{\arraystretch}{1.5}
\begin{tabular}{c c}
\hline
\hline
Parameter & Value (With $5\%$ variation)\\
\hline
Depth of the potential ($V_0$) & $-0.760 \pm 0.038$ $GeV$  \\
Potential strength ($\Lambda$)  & $1.079 \pm 0.054$ $GeV^{1.1}$ \\
Center of mass correction  ($E_{CM}$)  & $0.002 \pm 0.0001$ $GeV$ \\
$j-j$ coupling constant ($\sigma$) & $0.194 \pm 0.010
$ $GeV^{3}$\\
\hline
\hline
\end{tabular}
\endgroup
\end{table}

In the Independent Quark Model (IQM), all quarks are treated symmetrically and experience independent confinement. Consequently, the central potential parameters are allowed to vary for different heavy quarks. This treatment contrasts with the assumptions of Heavy Quark Effective Theory (HQET), where parameters are expected to remain identical; however, IQM does not invoke HQET.  

To derive the masses of the $P$, $D$, and $F$ states from the spin-averaged mass, three interactions are incorporated: spin-spin, spin-orbit, and tensor interactions. Based on phenomenological current confinement models for gluons, a closed analytical expression for the confined gluon propagator (CGP) in coordinate space was derived for low frequencies using a translationally invariant ansatz \cite{Vinodkumar:1992wu}. Utilizing the CGP, the confined one-gluon exchange potential (COGEP) between quarks has been formulated through the Fermi–Breit formalism. This formulation provides a consistent framework for investigating hadron spectroscopy and hadron–hadron interactions.  

The spin-orbit and tensor interaction terms appear as intrinsic components of the COGEP \cite{Vinodkumar:1992wu} and are treated as the sum of pairwise quark interactions, expressed as  
\begin{multline} 
    V^{LS}_{q_1 q_2 q_3}(r) = \frac{\alpha_s}{4} 
    \sum_{i=1,\,i<k}^{i,k=3} 
    \frac{N_{q_i}^2 N_{q_k}^2}
    {(E^D_{q_i} + m_{q_i})(E^D_{q_k} + m_{q_k})} 
    \frac{\lambda_i \cdot \lambda_j}{2r} \\
    \otimes 
    \Big[
    [r \times (\widehat{p}_{q_i} - \widehat{p}_{q_k}) \cdot (\widehat{\sigma}_{q_i} + \widehat{\sigma}_{q_k})]
    [D'_0(r) + 2D'_1(r)] \\
    + [r \times (\widehat{p}_{q_i} + \widehat{p}_{q_k}) \cdot (\widehat{\sigma}_{q_i} - \widehat{\sigma}_{q_k})]
    [D'_0(r) - D'_1(r)]
    \Big],
\end{multline}
\begin{multline}
    V^{T}_{q_1 q_2 q_3}(r) =  
    -\frac{\alpha_s}{4} 
    \sum_{i=1,\,i<k}^{i,k=3} 
    \frac{N_{q_i}^2 N_{q_k}^2}
    {(E^D_{q_i} + m_{q_i})(E^D_{q_k} + m_{q_k})} 
    \otimes \lambda_i \cdot \lambda_j \\
    \times 
    \left[
    \left(\frac{D''_1(r)}{3} - \frac{D'_1(r)}{3r}\right)
    S_{q_i q_k}
    \right],
\end{multline}
where $\lambda_i \cdot \lambda_j$ represents the color factor of the baryon, and 
$S_{q_i q_k} = [3(\sigma_{q_i} \cdot \hat{r})(\sigma_{q_k} \cdot \hat{r}) - \sigma_{q_i} \cdot \sigma_{q_k}]$.  

The running strong coupling constant $\alpha_s$ is defined as  
\begin{eqnarray} 
\alpha_s = 
\frac{\alpha_s(\mu_0)}
{1 + \frac{33 - 2n_f}{12\pi} \alpha_s(\mu_0) 
\ln\!\left(\frac{E_{q_1}^D + E_{q_2}^D + E_{q_3}^D}{\mu_0}\right)},
\end{eqnarray}  
where $\alpha_s(\mu_0 = 1\,\text{GeV}) = 0.6$ is adopted in the present work.  

The parametric forms of the confined gluon propagators $D_0(r)$ and $D_1(r)$ are retained as given in Ref.~\cite{Vinodkumar:1992wu}:  
\begin{equation} 
D_0(r) = 
\left(\frac{\alpha_1}{r} + \alpha_2\right)
\exp\!\left(\!-\frac{r^2 c_0^2}{2}\right),
\end{equation}
\begin{equation}
D_1(r) = 
\frac{\gamma}{r}
\exp\!\left(\!-\frac{r^2 c_1^2}{2}\right),
\end{equation}
with parameters 
$\alpha_1 = 0.036$, 
$\alpha_2 = 0.056$, 
$c_0 = 0.1017\ \text{GeV}$, 
$c_1 = 0.1522\ \text{GeV}$, 
and $\gamma = 0.0139$.  

While the gluon propagator in QCD is fundamentally flavor independent owing to the universal coupling of gluons to quark color charges, the phenomenological model represents confined gluon effects using fixed functional forms with system-dependent effective parameters $\alpha_1, \alpha_2, c_0, c_1, \gamma$. These parameters encapsulate nonperturbative QCD effects and hadron-specific dynamics. Their variation across different hadron sectors is consistent with the model’s phenomenological nature and supported by prior studies employing COGEP \cite{Shah:2014caa, Pandya:2021ddc, Shah:2014yma,Pandya:2023fak, Monteiro:2010}.

After obtaining the wavefunction, the normalization constant $N_{q_i}$ for each individual quark can be determined. This normalization allows the evaluation of the matrix elements $\langle \psi | V^{LS} | \psi \rangle$ and $\langle \psi | V^{T} | \psi \rangle$ for all possible permutations of $q_1$, $q_2$, and $q_3$. The total contribution for a given state is then obtained by summing over all these permutations. By including the spin-spin interaction terms in this total, the masses of the corresponding $P$, $D$, and $F$ states can be calculated.  

Our predicted masses, along with the available experimental data and other theoretical predictions, are presented in Tables~\ref{sc}, \ref{pc}, \ref{dc}, and \ref{fc} for the $S$, $P$, $D$, and $F$ states of the $\Xi_c^0$ baryon, and in Tables~\ref{sb}, \ref{pb}, \ref{db}, and \ref{fb} for the corresponding states of the $\Xi_b^-$ baryon respectively.

\begin{table*}[ht]
{
\begingroup
\caption{$S$ State masses of $\Xi_c^{0}$(in $GeV$)} \label{sc}
\setlength{\tabcolsep}{5pt}
\renewcommand{\arraystretch}{1.9}
\begin{tabular}{c c c c c c c c c c}
\\
\hline
\hline
$nL$ & $J^{P}$ & State & $\big<V^{jj}_{q_1q_2q_3}\big>$ & Our & Exp.\cite{ParticleDataGroup:2024cfk} &\cite{Brown:2014ena} & \cite{Shah:2016mig} & \cite{Gandhi:2019bju} & \cite{Ebert:2011kk} \\
\hline
$1S$ & $\frac{1}{2}^+$ & $1^2S_{\frac{1}{2}}$ & $-0.111$ & $2.475 \pm 0.044$ & $2.470\pm0.0003$ & $2.433$  & $2.470 $ & $2.471$ & $2.476$\\ 

$1S$ & $\frac{3}{2}^+$ & $1^4S_{\frac{3}{2}}$ & $0.066$ & $2.652\pm0.042$ & $2.646\pm0.0002$ &$2.648$  & $2.585$ & $2.648$ & $-$\\

$2S$ & $\frac{1}{2}^+$ & $2^2S_{\frac{1}{2}}$ & $-0.081$ & $2.871 \pm0.062$ & $-$ & $-$ & $2.831$ & $2.964$ & $2.959$  \\
$2S$ & $\frac{3}{2}^+$ & $2^4S_{\frac{3}{2}}$ & $0.049$ & $3.001\pm0.059$ & $-$  &$-$ & $2.919$ & $3.080$ & $-$ \\

$3S$ & $\frac{1}{2}^+$ & $3^2S_{\frac{1}{2}}$ & $-0.070$ & $3.096\pm0.070$ & $-$  & $-$ & $3.097$ & $3.358$ & $3.323$\\
$3S$ & $\frac{3}{2}^+$ & $3^4S_{\frac{3}{2}}$ & $0.042$ & $3.207\pm0.068$ & $-$  & $-$ & $3.151$ & $3.424$ & $-$ \\

$4S$ & $\frac{1}{2}^+$ & $4^2S_{\frac{1}{2}}$ & $-0.063$ & $3.256 \pm0.077$ & $-$& $-$ & $3.339$ & $3.720$ & $3.632$\\

$4S$ & $\frac{3}{2}^+$ & $4^4S_{\frac{3}{2}}$ & $0.038$ & $3.356\pm0.075$ & $-$& $-$  & $3.375$ & $3.763$& $-$\\

$5S$ & $\frac{1}{2}^+$ & $5^2S_{\frac{1}{2}}$ & $-0.058$ & $3.380\pm0.082$ & $-$& $-$ & $3.566$ & $4.064$ & $3.909$ \\

$5S$ & $\frac{3}{2}^+$ & $5^4S_{\frac{3}{2}}$ & $0.035$ & $3.473\pm0.081$& $-$& $-$ & $3.591$ & $4.093$ & $-$\\
\hline
\hline
\end{tabular}
\endgroup}
\end{table*}

\begin{table*}[ht]
{
\begingroup
\caption{$P$ State masses $\Xi^0_c$(in $GeV$)} \label{pc}
\setlength{\tabcolsep}{9pt}
\renewcommand{\arraystretch}{1.5}
\begin{tabular}{ c c c c c c c c c c}
\hline
\hline
$n^{2S+1}L_J$ & $\big<V^{jj}_{q_1q_2q_3}\big>$& $\big<V_{q_1q_2q_3}^{L.S}\big>$ & $\big<V_{q_1q_2q_3}^T\big>$& Our & Exp.\cite{ParticleDataGroup:2024cfk} & \cite{Shah:2016mig} & \cite{Gandhi:2019bju} & \cite{Ebert:2011kk} \\
\hline
$1^2P_{\frac{1}{2}}$ & $-0.096$ &$-0.113$ & $-0.080$ & $2.561\pm0.072$ & $-$  & $2.709$ & $2.828$ & $2.792$\\
$1^2P_{\frac{3}{2}}$ &$0.062$ &$-0.019$ & $0.003$ & $2.896\pm0.052$ & $2.882\pm0.009$ & $2.707$ & $2.820$ & $2.819$\\
$1^4P_{\frac{1}{2}}$ & $ -0.275$ & $-0.160$ & $-0.156$ & $2.260\pm0.090$ & $-$ & $2.710$ & $2.832$ & $-$\\
$1^4P_{\frac{3}{2}}$ & $-0.106$ & $-0.066$ & $0.072$ &$2.750\pm0.030$ & $-$   & $2.708$ & $2.824$ & $-$\\
$1^4P_{\frac{5}{2}}$ & $0.141$ & $0.058$ & $-0.015$ & $3.035\pm0.044$ & $-$   & $2.705$ & $2.814$& $-$\\
\hline
$2^2P_{\frac{1}{2}}$ & $-0.076$ & $-0.063$ & $-0.036$ & $2.919\pm0.077$ & $2.923\pm0.0004$   & $2.967$ & $3.191$ & $3.179$\\
$2^2P_{\frac{3}{2}}$ & $0.047$ & $-0.011$ & $0.001$ & $3.132\pm0.064$ & $-$   & $2.964$ & $3.184$ & $3.201$\\
$2^4P_{\frac{1}{2}}$ & $-0.215$ & $-0.090$ & $-0.071$ & $2.719\pm0.089$ & $-$   & $2.968$ & $3.195$ & $-$\\
$2^4P_{\frac{3}{2}}$ & $-0.080$ & $-0.037$ & $0.033$ & $3.010\pm0.070$ & $-$  & $2.966$ & $3.188$ & $-$\\
$2^4P_{\frac{5}{2}}$ & $0.109$ & $0.033$ & $-0.007$ & $3.230\pm0.058$  & $-$  & $2.962$ & $3.177$ & $-$\\
\hline
$3^2P_{\frac{1}{2}}$ & $-0.067$ & $-0.039$ & $-0.017$ & $3.140\pm0.081$   & $-$ & $3.205$ & $3.541$ & $3.5$\\
$3^2P_{\frac{3}{2}}$ & $0.041$ & $-0.007$ & $0.001$ & $3.298\pm0.0714$   & $-$ & $3.203$ & $3.533$ & $3.519$\\
$3^4P_{\frac{1}{2}}$ & $-0.188$ & $-0.056$ & $-0.034$ & $2.985\pm0.090$   & $-$ & $3.207$ & $3.545$ & $-$\\
$3^4P_{\frac{3}{2}}$ & $-0.069$ & $-0.023$ & $0.012$ & $3.187\pm0.078$   & $-$ & $3.204$ & $3.537$ & $-$\\
$3^4P_{\frac{5}{2}}$ & $0.095$ & $0.020$ & $-0.003$ & $3.375\pm0.067$   & $-$ & $3.200$ & $3.527$ & $-$\\
\hline
$4^2P_{\frac{1}{2}}$ & $-0.061$ & $-0.026$ & $-0.009$ & $3.296\pm0.085$   & $-$ & $3.431$ & $3.879$ & $3.785$\\
$4^2P_{\frac{3}{2}}$ & $0.037$ & $-0.004$ & $0.000$ & $3.426\pm0.077$   & $-$ & $3.429$ & $3.871$ & $3.804$\\
$4^4P_{\frac{1}{2}}$ & $-0.171$ & $-0.037$ & $-0.018$ & $3.166\pm0.093$   & $-$ & $3.433$ & $3.883$ & $-$\\
$4^4P_{\frac{3}{2}}$ & $-0.062$ & $-0.015$ & $0.008$ & $3.324\pm0.084$   & $-$ & $3.430$ & $3.875$ & $-$\\
$4^4P_{\frac{5}{2}}$ & $0.086$ & $0.014$ & $-0.002$ & $3.491\pm0.073$   & $-$ & $3.426$ & $3.865$ & $-$\\
\hline
\hline
\end{tabular} 
\endgroup}
\end{table*}

\begin{table*}[ht]
{
\begingroup
\caption{$D$ State masses $\Xi_c^0$ (in $GeV$)} \label{dc}
\setlength{\tabcolsep}{6pt}
\renewcommand{\arraystretch}{1.5}
\begin{tabular}{ c c c c c c c c c c c c}
\hline
\hline
$n^{2S+1}L_J$ & $\big<V^{jj}_{q_1q_2q_3}\big>$& $\big<V^{L.S}_{q_1q_2q_3}\big>$ & $\big<V^{T}_{q_1q_2q_3}\big>$& Our &Exp.\cite{ParticleDataGroup:2024cfk}& \cite{Shah:2016mig} & \cite{Gandhi:2019bju} & \cite{Ebert:2011kk} \\
\hline
$1^2D_{\frac{3}{2}}$ & $-0.178$ & $-0.081$ & $-0.011$ &$2.755\pm0.070$ & $-$ &  $2.927$ & $3.116$ & $-$\\
$1^2D_{\frac{5}{2}}$ & $-0.029$ & $-0.002$ & $0.001$ & $2.993\pm0.063$& $-$ &  $2.927$ & $3.103$ & $-$\\
$1^4D_{\frac{1}{2}}$ & $-0.150$ &$-0.169$ & $-0.051$ &$2.655\pm0.078$ &$-$ & $2.928$ & $3.131$ & $-$\\
$1^4D_{\frac{3}{2}}$ & $-0.064$ & $-0.110$ & $-0.017$ & $2.832\pm0.071$ &$-$ & $2.923$ & $3.121$ & $3.059$\\
$1^4D_{\frac{5}{2}}$ & $0.133$ & $-0.032$ & $0.005$ & $3.130\pm0.063$ &$-$ & $2.924$ & $3.108$ & $3.076$\\
$1^4D_{\frac{7}{2}}$ & $0.207$ & $0.069$ & $-0.004$ & $3.296\pm0.057$ &$-$ & $2.919$ & $3.092$ & $-$\\
\hline 
$2^2D_{\frac{3}{2}}$ & $-0.138$ & $-0.050$ & $-0.006$ & $3.015\pm0.076$ & $-$ & $3.169$ & $3.464$ & $-$\\
$2^2D_{\frac{5}{2}}$ & $-0.019$ & $-0.002$ & $0.001$ & $3.189\pm0.071$ & $-$ & $3.165$ & $3.452$ & $-$\\
$2^4D_{\frac{1}{2}}$ & $-0.121$ & $-0.105$ & $-0.027$ & $2.956\pm0.081$& $2.966\pm0.002$ & $3.166$ & $3.478$ & $-$\\
$2^4D_{\frac{3}{2}}$ & $-0.053$ & $-0.069$ & $-0.009$ & $3.078\pm0.076$ & $3.080\pm0.001$ & $3.161$ & $3.469$ & $3.388$\\
$2^4D_{\frac{5}{2}}$ & $0.105$ & $-0.020$ & $0.002$ & $3.297\pm0.071$ & $-$ & $3.162$ & $3.457$ & $3.407$\\
$2^4D_{\frac{7}{2}}$ & $0.165$ & $0.043$ & $-0.002$ & $3.415\pm0.066$ & $-$ & $3.157$ & $3.442$ & $-$\\
\hline
$3^2D_{\frac{3}{2}}$ & $-0.120$ & $-0.034$ & $-0.003$ & $3.191\pm0.081$& $-$ & $-$& $3.804$ & $-$\\
$3^2D_{\frac{5}{2}}$ & $-0.016$ & $-0.001$ & $0.000$ & $3.332\pm0.077$& $-$& $-$ & $3.792$ & $-$\\
$3^4D_{\frac{1}{2}}$ & $-0.107$ & $-0.071$ & $ -0.015$ & $3.156\pm0.084$& $-$ & $-$& $3.817$ & $-$\\
$3^4D_{\frac{3}{2}}$ & $-0.047$ & $-0.046$ & $-0.005$ & $3.250\pm0.081$ & $-$& $-$ & $3.308$ & $3.678$\\
$3^4D_{\frac{5}{2}}$ & $0.092$ &$-0.013$ & $0.001$ & $3.430\pm0.080$ & $-$ & $-$& $3.796$ & $3.699$\\
$3^4D_{\frac{7}{2}}$ & $0.145$ & $0.029$ & $-0.001$ & $3.521\pm0.074$& $-$ & $-$& $3.782$ & $-$\\
\hline
$4^2D_{\frac{3}{2}}$ & $-0.110$ & $-0.024$ & $-0.002$ & $3.326\pm0.085$& $-$ & $-$& $4.132$ & $-$\\
$4^2D_{\frac{5}{2}}$ & $-0.014$ & $-0.001$ & $0.000$ & $3.447\pm0.082$& $-$& $-$ & $4.121$ & $-$\\
$4^4D_{\frac{1}{2}}$ & $-0.098$ & $-0.050$ & $-0.009$ & $3.305\pm0.088$& $-$ & $-$& $4.144$ & $-$\\
$4^4D_{\frac{3}{2}}$ & $-0.043$ & $-0.032$ & $-0.003$ & $3.382\pm0.085$ & $-$ & $-$& $4.136$ & $3.945$\\
$4^4D_{\frac{5}{2}}$ & $0.084$ & $-0.009$ & $0.001$ & $3.537\pm0.082$& $-$& $-$ & $4.125$ & $3.965$\\
$4^4D_{\frac{7}{2}}$ & $0.133$ & $0.020$ & $-0.001$ & $3.613\pm0.079$& $-$& $-$ & $4.112$ & $-$\\
\hline
\hline
\end{tabular}
\endgroup}
\end{table*}

\begin{table*}
{
\begingroup
\caption{$F$ State masses $\Xi_c^0$ (in $GeV$)} \label{fc}
\setlength{\tabcolsep}{9pt}
\renewcommand{\arraystretch}{1.5}
\begin{tabular}{c c c c c c c c c}

\hline
\hline
$n^{2S+1}L_J$ & $\big<V^{jj}_{q_1q_2q_3}\big>$& $\big<V^{L.S}_{q_1q_2q_3}\big>$ & $\big<V^{T}_{q_1q_2q_3}\big>$& Our & Exp.\cite{ParticleDataGroup:2024cfk}& \cite{Shah:2016mig} & \cite{Gandhi:2019bju} & \cite{Ebert:2011kk}\\
\hline
$1^2F_{\frac{5}{2}}$ & $-0.096$ & $-0.066$ & $-0.003$ & $2.989\pm0.073$ & $-$ & $3.124$ & $3.388$ & $3.278$\\
$1^2F_{\frac{7}{2}}$ & $-0.097$ & $0.005$ & $0.000$ & $3.063\pm0.069$ &$-$ & $3.118$ & $3.369$ & $-$\\
$1^4F_{\frac{3}{2}}$ & $-0.369$ & $-0.144$ & $-0.017$ & $2.625\pm0.082$ &$-$& $3.131$ & $3.408$ & $-$\\
$1^4F_{\frac{5}{2}}$ & $-0.512$ & $-0.087$ & $-0.004$ & $2.551\pm0.079$ & $-$ & $3.126$ & $3.393$ & $-$\\
$1^4F_{\frac{7}{2}}$ & $0.398$ & $-0.016$ & $0.003$ & $3.539\pm0.066$ & $-$ & $3.120$ & $3.375$ & $3.292$\\
$1^4F_{\frac{9}{2}}$ & $0.539$ & $0.069$ & $-0.002$ & $3.760\pm0.060$ & $-$ & $3.113$ & $3.358$ & $-$\\
\hline
$2^2F_{\frac{5}{2}}$ & $-0.073$ & $-0.043$ & $-0.002$ & $3.186\pm0.079$ & $-$ &$-$ & $3.727$ & $3.575$\\
$2^2F_{\frac{7}{2}}$ & $-0.074$ & $0.003$ & $0.000$ & $3.234\pm0.076$ &$-$ &$-$ & $3.710$ & $-$\\
$2^4F_{\frac{3}{2}}$ & $-0.303$ & $-0.094$ & $-0.010$ & $2.897\pm0.085$  &$-$ &$-$& $3.745$ & $-$\\
$2^4F_{\frac{5}{2}}$ & $-0.420$ & $-0.057$ & $-0.002$ & $2.825\pm0.084$  &$-$ &$-$& $3.732$ & $-$\\
$2^4F_{\frac{7}{2}}$ & $0.320$ & $-0.010$ & $0.002$ & $3.616\pm0.072$  &$-$ &$-$ & $3.715$ & $3.592$\\
$2^4F_{\frac{9}{2}}$ & $0.435$ & $0.045$ & $-0.001$ & $3.783\pm0.068$  &$-$ &$-$& $3.695$ & $-$\\
\hline
$3^2F_{\frac{5}{2}}$ & $-0.063$ & $-0.030$ & $-0.001$ & $3.329\pm0.083$& $-$ &$-$ & $4.055$ & $3.845$\\
$3^2F_{\frac{7}{2}}$ & $-0.064$ & $0.002$ & $0.002$ & $3.362\pm0.081$& $-$ &$-$ & $4.042$ & $-$\\
$3^4F_{\frac{3}{2}}$ & $-0.271$ & $-0.065$ & $-0.006$ & $3.081\pm0.088$ &$-$ &$-$ & $4.069$ & $-$\\
$3^4F_{\frac{5}{2}}$ & $-0.375$ & $-0.039$ & $-0.001$ & $3.007\pm0.088$& $-$ &$-$ & $4.059$ & $-$\\
$3^4F_{\frac{7}{2}}$ & $0.284$ & $-0.007$ & $0.001$ & $3.701\pm0.078$ & $-$ &$-$ & $4.046$ & $3.865$\\
$3^4F_{\frac{9}{2}}$ & $0.386$ & $0.031$ & $-0.001$ & $3.840\pm0.074$ &$-$ &$-$ & $4.030$ & $-$\\
\hline
\end{tabular}
\endgroup}
\end{table*}

\begin{table*}[ht]
{
\begingroup
\caption{$S$ State masses of $\Xi_b^{-}$(in $GeV$)} \label{sb}
\setlength{\tabcolsep}{9pt}
\renewcommand{\arraystretch}{1.5}
\begin{tabular}{c c c c c c c  c c c}
\hline
\hline
$nL$ & $J^{P}$ & State & $\big<V^{jj}_{q_1q_2q_3}\big>$ & Our& Exp.\cite{ParticleDataGroup:2024cfk} & Lattice QCD\cite{Brown:2014ena} & \cite{Oudichhya:2021yln}  & \cite{Ebert:2011kk}\\
\hline
$1S$ & $\frac{1}{2}^+$ & $1^2S_{\frac{1}{2}}$ &$-0.099$ & $5.804\pm0.068$& $5.797\pm0.0004$ & $5.771$ &$5.797$  & $5.803$\\
$1S$ & $\frac{3}{2}^+$ & $1^4S_{\frac{3}{2}}$ & $0.060$ & $5.962\pm0.066$ &$5.956\pm0.0004$ & $5.960$ &$5.955$  & $-$ \\

$2S$ & $\frac{1}{2}^+$ & $2^2S_{\frac{1}{2}}$ & $-0.077$ & $6.119\pm0.082$ & $-$ & $-$ & $6.189$ &  $6.266$\\
$2S$ & $\frac{3}{2}^+$ & $2^4S_{\frac{3}{2}}$ & $0.046$ & $6.242\pm0.079$ &$-$ & $-$ & $6.298$ & $-$\\

$3S$ & $\frac{1}{2}^+$ & $3^2S_{\frac{1}{2}}$ & $   -0.067$ & $6.299\pm0.089$ & $-$ & $-$ & $6.558$ & $6.601$\\
$3S$ & $\frac{3}{2}^+$ & $3^4S_{\frac{3}{2}}$ & $0.040$ &$6.407\pm0.087$  & $-$& $-$ & $6.623$  &$-$\\

$4S$ & $\frac{1}{2}^+$ & $4^2S_{\frac{1}{2}}$ & $-0.062$ & $6.427\pm0.095$ & $-$ & $-$ & $6.907$ & $6.913$\\
$4S$ & $\frac{3}{2}^+$ & $4^4S_{\frac{3}{2}}$ & $0.037$ &$6.526\pm0.093$ & $-$ & $-$ & $6.933$&$-$ \\

$5S$ & $\frac{1}{2}^+$ & $5^2S_{\frac{1}{2}}$ & $-0.058$ &$6.526\pm0.099$ & $-$& $-$  & $7.239$ & $7.165$\\
$5S$ & $\frac{3}{2}^+$ & $5^4S_{\frac{3}{2}}$ & $0.035$ & $6.619\pm0.097$ & $-$ & $-$ & $7.230$  &$-$\\
\hline
\hline
\end{tabular}
\endgroup}
\end{table*}

\begin{table*}[ht]
{
\begingroup
\caption{$P$ State masses of $\Xi^-_b$(in $GeV$)} \label{pb}
\setlength{\tabcolsep}{9pt}
\renewcommand{\arraystretch}{1.5}
\begin{tabular}{ c c c c c c c c}
\hline
\hline
$n^{2S+1}L_J$ & $\big<V^{jj}_{q_1q_2q_3}\big>$& $\big<V_{q_1q_2q_3}^{L.S}\big>$ & $\big<V_{q_1q_2q_3}^T\big>$& Our & Exp.\cite{ParticleDataGroup:2024cfk} & \cite{Oudichhya:2021yln} & \cite{Ebert:2011kk}\\
\hline
$1^2P_{\frac{1}{2}}$ & $-0.088$ & $-0.089$ & $-0.063$ & $5.874\pm0.091$ & $-$ & $-$ & $6.120$\\
$1^2P_{\frac{3}{2}}$ & $0.056$ & $-0.015$ & $0.002$ & $6.158\pm0.074$ & $-$ & $6.098$ & $6.130$\\
$1^4P_{\frac{1}{2}}$ & $-0.251$ & $-0.126$ & $-0.123$ & $5.614\pm0.108$ & $-$& $-$ & $-$\\
$1^4P_{\frac{3}{2}}$ & $-0.096$ & $-0.052$ &$0.057$ & $6.023\pm0.082$ & $-$& $-$& $-$\\
$1^4P_{\frac{5}{2}}$ & $0.128$ & $0.046$ & $-0.012$ & $6.277\pm0.066$& $-$ & $6.243$& $-$\\
\hline
$2^2P_{\frac{1}{2}}$ & $-0.073$ & $-0.053$ & $-0.030$ & $6.153\pm0.095$ & $-$& $-$ & $6.496$\\
$2^2P_{\frac{3}{2}}$ & $0.045$ & $-0.009$ & $0.001$ & $6.346\pm0.083$ & $-$ & $6.437$ & $6.502$\\
$2^4P_{\frac{1}{2}}$ & $-0.206$ & $-0.076$ & $-0.060$ & $5.969\pm0.107$ & $-$ & $-$& $-$\\
$2^4P_{\frac{3}{2}}$ & $-0.076$ & $-0.031$ & $0.028$ & $6.230\pm0.090$ & $-$ & $-$& $-$\\
$2^4P_{\frac{5}{2}}$ & $0.104$ & $0.028$ & $-0.006$ & $6.435\pm0.077$ & $-$ & $6.428$ & $-$\\
\hline
$3^2P_{\frac{1}{2}}$ &  $-0.065$ & $-0.034$ & $-0.015$ & $6.330\pm0.099$  & $-$& $-$& $6.805$\\
$3^2P_{\frac{3}{2}}$ & $0.040$ & $-0.006$ & $0.000$ & $6.479\pm0.089$ & $-$ & $6.759$  & $6.810$\\
$3^4P_{\frac{1}{2}}$ & $-0.183$ & $-0.049$ & $-0.030$ & $6.183\pm0.108$  & $-$& $-$& $-$\\
$3^4P_{\frac{3}{2}}$ &  $-0.067$ & $-0.020$ & $0.014$ &$6.371\pm0.096$ & $-$& $-$& $-$\\
$3^4P_{\frac{5}{2}}$ & $0.092$ & $0.018$ & $-0.003$ & $6.552\pm0.085$ & $-$ & $6.608$ & $-$\\
\hline
$4^2P_{\frac{1}{2}}$ & $-0.058$ & $-0.024$ & $-0.008$ & $6.543\pm0.015$  & $-$& $-$&$7.068$\\
$4^2P_{\frac{3}{2}}$ & $0.036$ & $-0.004$ & $0.000$ & $6.665\pm0.010$ & $-$ & $7.066$ & $7.073$\\
$4^4P_{\frac{1}{2}}$ & $-0.164$ & $-0.033$ & $-0.016$ & $6.419\pm0.020$  & $-$& $-$& $-$\\
$4^4P_{\frac{3}{2}}$ & $-0.060$ & $-0.014$ & $0.007$ & $6.566\pm0.014$  & $-$& $-$& $-$\\
$4^4P_{\frac{5}{2}}$ & $0.083$ & $0.012$ & $-0.002$ & $6.726\pm0.008$ & $-$ & $6.783$  & $-$\\
\hline
\hline
\end{tabular}
\endgroup}
\end{table*}

\begin{table*}[ht]
{
\begingroup
\caption{$D$ State masses of $\Xi_b^-$ (in $GeV$)} \label{db}
\setlength{\tabcolsep}{9pt}
\renewcommand{\arraystretch}{1.5}
\begin{tabular}{ c c c c c c c c }
\hline
\hline
$n^{2S+1}L_J$ & $\big<V^{jj}_{q_1q_2q_3}\big>$& $\big<V^{L.S}_{q_1q_2q_3}\big>$ & $\big<V^{T}_{q_1q_2q_3}\big>$& Our &Exp.\cite{ParticleDataGroup:2024cfk}&\cite{Oudichhya:2021yln} & \cite{Ebert:2011kk}\\
\hline
$1^2D_{\frac{3}{2}}$ & $-0.162$ & $-0.065$ & $-0.009$ & $6.018\pm0.090$& $-$ & $-$ & $-$ \\
$1^2D_{\frac{5}{2}}$ & $-0.025$ & $-0.002$ & $0.001$ & $6.227\pm0.083$& $6.228\pm0.001$ & $6.385$ & $-$  \\
$1^4D_{\frac{1}{2}}$ & $-0.146$ & $-0.136$ & $-0.041$ & $5.931\pm0.096$ & $-$ & $-$ & $-$ \\
$1^4D_{\frac{3}{2}}$ & $-0.060$ & $-0.088$ & $-0.014$ & $6.091\pm0.090$& $-$ & $-$ & $6.366$  \\
$1^4D_{\frac{5}{2}}$ & $0.122$ & $-0.026$ & $0.004$ & $6.354\pm0.082$ & $-$ & $-$ & $6.373$ \\
$1^4D_{\frac{7}{2}}$ & $0.191$ & $0.056$ & $-0.004$ & $6.496\pm0.076$ & $-$ & $6.518$ & $-$ \\
\hline
$2^2D_{\frac{3}{2}}$ & $-0.132$ & $-0.031$ & $-0.003$ & $6.235\pm0.094$ & $-$  & $-$ & $-$ \\
$2^2D_{\frac{5}{2}}$ & $-0.018$ & $-0.001$ & $0.000$ &$6.383\pm0.090$ &$-$& $6.696$  & $-$ \\
$2^4D_{\frac{1}{2}}$ & $-0.122$ & $-0.066$ & $-0.013$ & $6.200\pm0.097$ & $-$ & $-$ & $-$ \\
$2^4D_{\frac{3}{2}}$ & $-0.051$ & $-0.043$ & $-0.004$ & $6.303\pm0.093$ & $-$ & $-$ & $6.690$ \\
$2^4D_{\frac{5}{2}}$ & $0.101$ & $-0.012$ & $0.001$ & $6.491\pm0.088$ & $-$ & $-$ & $6.696$ \\
$2^4D_{\frac{7}{2}}$ & $0.158$ & $0.026$ & $-0.001$ & $6.584\pm0.085$ & $-$ & $6.661$ & $-$ \\
\hline
$3^2D_{\frac{3}{2}}$ & $-0.117$ & $-0.021$ & $-0.002$ & $6.372\pm0.098$ & $-$  & $-$ & $-$ \\
$3^2D_{\frac{5}{2}}$ & $-0.015$ & $-0.001$ & $0.000$ & $6.497\pm0.095$ & $-$ & $6.993$& $-$ \\
$3^4D_{\frac{1}{2}}$ & $-0.110$ & $-0.044$ & $-0.007$ & $6.351\pm0.100$ & $-$ & $-$ & $-$ \\
$3^4D_{\frac{3}{2}}$ & $-0.046$ & $-0.029$ & $-0.002$ & $6.435\pm0.097 $ & $-$ & $-$ & $6.966$ \\
$3^4D_{\frac{5}{2}}$ & $0.090$ & $-0.008$ & $0.001$ & $6.595\pm0.093$ & $-$ & $-$ & $6.970$ \\
$3^4D_{\frac{7}{2}}$ & $0.141$ & $0.018$ & $-0.001$ & $6.672\pm0.091$& $-$ & $6.801$ & $-$ \\
\hline
$4^2D_{\frac{3}{2}}$ & $-0.108$ & $-0.014$ & $-0.001$ & $6.479\pm0.102$ & $-$ & $-$ & $-$ \\
$4^2D_{\frac{5}{2}}$ & $-0.013$ & $-0.000$ & $0.000$ & $6.589\pm0.099$ & $-$ & $7.278$ & $-$ \\
$4^4D_{\frac{1}{2}}$ & $-0.102$ & $-0.030$ & $-0.004$ & $6.466\pm0.103$ & $-$ & $-$ & $-$ \\
$4^4D_{\frac{3}{2}}$ & $-0.043$ & $-0.019$ & $-0.002$ & $6.538\pm0.101$ & $-$ & $-$ & $7.208$ \\
$4^4D_{\frac{5}{2}}$ & $0.084$ & $-0.006$ & $0.000$ & $6.681\pm0.097$ & $-$ & $-$ & $7.212$ \\
$4^4D_{\frac{7}{2}}$ & $0.131$ & $0.012$ & $-0.000$ & $6.745\pm0.096$ & $-$ & $6.938$ & $-$ \\
\hline
\hline
\end{tabular}
\endgroup}  
\end{table*}

\begin{table*}
{
\begingroup
\caption{$F$ State masses of $\Xi^-_b$ (in $GeV$)} \label{fb}
\setlength{\tabcolsep}{9pt}
\renewcommand{\arraystretch}{1.5}
\begin{tabular}{c c c c c c c c}

\hline
\hline
$n^{2S+1}L_J$ & $\big<V^{jj}_{q_1q_2q_3}\big>$& $\big<V^{L.S}_{q_1q_2q_3}\big>$ & $\big<V^{T}_{q_1q_2q_3}\big>$& Our & Exp.\cite{ParticleDataGroup:2024cfk}& \cite{Oudichhya:2021yln} & \cite{Ebert:2011kk}\\
\hline
$1^2F_{\frac{5}{2}}$ & $-0.087$ & $-0.053$ & $-0.003$ & $6.214\pm0.092$ &$-$&$-$&$6.577$ \\
$1^2F_{\frac{7}{2}}$ & $-0.088$ & $0.004$ & $0.000$ & $6.274\pm0.089$ &$-$ &$6.659$&$-$\\
$1^4F_{\frac{3}{2}}$ & $-0.343$ & $-0.117$ & $-0.014$ & $5.883\pm0.101$ &$-$ &$-$&$-$\\
$1^4F_{\frac{5}{2}}$ & $-0.478$ & $-0.071$ & $-0.003$ & $5.806\pm0.100$ &$-$&$-$&$-$\\
$1^4F_{\frac{7}{2}}$ & $0.368$ & $-0.013$ & $0.002$ & $6.715\pm0.082$ &$-$&$-$&$6.581$\\
$1^4F_{\frac{9}{2}}$ & $0.499$ & $0.056$ & $-0.002$ & $6.911\pm0.076$ &$-$&$6.782$&$-$\\
\hline
$2^2F_{\frac{5}{2}}$ & $-0.069$ & $-0.037$ & $-0.002$ & $6.369\pm0.096$ &$-$&$-$&$6.863$\\
$2^2F_{\frac{7}{2}}$ & $-0.070$ & $0.003$ & $0.000$ & $6.410\pm0.094$ &$-$&$-$&$-$\\
$2^4F_{\frac{3}{2}}$ & $-0.292$ & $-0.081$ & $-0.008$ & $6.095\pm0.104$ &$6.100\pm0.0004$&$-$&$-$\\
$2^4F_{\frac{5}{2}}$ & $-0.405$ & $-0.049$ & $-0.002$ & $6.021\pm0.103$ &$-$&$-$&$-$\\
$2^4F_{\frac{7}{2}}$ & $0.308$ & $-0.009$ & $0.001$ & $6.777\pm0.088$ &$-$&$-$&$6.867$\\
$2^4F_{\frac{9}{2}}$ & $0.418$ & $0.039$ & $-0.001$ & $6.933\pm0.083$ &$-$&$-$&$-$\\
\hline
$3^2F_{\frac{5}{2}}$ & $-0.061$ & $-0.026$ & $-0.001$ & $6.483\pm0.100$ &$-$&$-$&$7.114$\\
$3^2F_{\frac{7}{2}}$ & $-0.062$ & $0.002$ & $0.000$ & $6.512\pm0.098$ &$-$&$-$&$-$\\
$3^4F_{\frac{3}{2}}$ & $-0.26$ & $-0.058$ & $-0.005$ & $6.243\pm0.106$ &$-$&$-$&$-$\\
$3^4F_{\frac{5}{2}}$ & $-0.368$ & $-0.035$ & $-0.001$ & $6.168\pm0.106$ &$-$&$-$&$-$\\
$3^4F_{\frac{7}{2}}$ & $0.278$ & $-0.006$ & $0.001$ & $6.844\pm0.092$ &$-$&$-$&$7.117$\\
$3^4F_{\frac{9}{2}}$ & $0.378$ & $0.028$ & $-0.001$ & $6.977\pm0.089$ &$-$&$-$&$-$\\
\hline

\end{tabular}
\endgroup}
\end{table*}

\section{Magnetic moments and Decay properties}\label{sec:3}
\subsection{Magnetic Moment}
The magnetic moment of baryons is expressed in terms of constituent quarks and spin-flavor wave function\cite{Patel:2007gx} as given below:
\begin{eqnarray}
    \mu_{B} = \sum_{q}\bigg<\phi_{sf}|\vec{\mu}_{qz}|\phi_{sf}\bigg>, 
    \end{eqnarray}
    where
    \begin{eqnarray}\label{m}
    \mu_{q} = \frac{e_{q}}{2 m_{q}}\sigma_{q}. 
    \end{eqnarray}
Here, $e_q$ and $\sigma_q$ represent the charge and the spin of the quark, and $|\phi_{sf}\big>$ is the spin-flavor wave function. The spin-flavor wave functions of $\Xi_c^0$ \& $\Xi_b^-$ baryons with $J^P=1/2^+$ are
\begin{multline}
    |\phi_{sf}\big>_{\Xi_c^0} = \frac{-1}{6}(d_+s_-c_+ + s_+d_-c_+ + c_+s_-d_+ \\+ c_+d_-s_+ - 2d_+c_-s_+ - 2s_+c_-d_+ +\\d_-s_+c_+ + s_-d_+c_+ - 2c_-s_+d_+\\ - 2c_-d_+s_+ + d_-c_+s_+ + s_-c_+d_+ \\-2d_+s_+c_- - 2s_+d_+c_- +c_+s_+d_- \\ + c_+d_+s_- + d_+c_+s_- + s_+c_+d_-)
\end{multline}
\begin{multline}
    |\phi_{sf}\big>_{\Xi_b^-} = \frac{-1}{6}(d_+s_-b_+ + s_+d_-b_+ + b_+s_-d_+ \\+b_+d_-s_+ - 2d_+b_-s_+ - 2s_+ b_-d_+ +\\d_-s_+b_+ + s_- d_+b_+ -2b_-s_+d_+ \\-2b_-d_+s_+ +d_-b_+s_+ + s_-b_+d_+\\-2d_+s_+b_- - 2s_+d_+b_- + b_+s_+d_- \\+b_+d_+s_- + d_+b_+s_- + s_+b_+d_-)
\end{multline}
and the spin-flavor wave functions of $\Xi_c^0$ \& $\Xi_b^-$ baryons with $J^P=3/2^+$ are
\begin{multline}
    |\phi_{sf}\big>_{\Xi_c^0} = \frac{1}{\sqrt{3}}(d_+d_+c_+ + d_+c_+d_+ +c_+d_+d_+),
\end{multline}
\begin{multline}
    |\phi_{sf}\big>_{\Xi_b^-} = \frac{1}{\sqrt{6}}(d_+s_+b_+ +s_+d_+b_+ + \\d_+b_+s_+ + b_+d_+s_+ + s_+b_+d_+ + b_+s_+d_+).
\end{multline}
Here we have used the notations for spin operator $\hat{S}|q\uparrow\big>=\hat{S}|q+\big>$ \& $\hat{S}|q\downarrow\big>=\hat{S}|q-\big>$.
The masses of quarks inside a baryon undergo alteration due to the interquark interaction present in the bound system. This modified mass is referred to as the effective quark mass, in our model the effective quark masses $m_{q}^{eff}$ is defined as
\begin{eqnarray} 
    m_{q}^{eff} = E^D_q\left(1+\frac{\big<H\big>-E_{CM}}{\sum_q E^D_q}\right).
\end{eqnarray}
Here, the $\big<H\big>$ includes the strength of spin-spin interactions only as we are calculating the magnetic field of $S$ waves.
The effective quark mass follows the property of $M_J = \sum_{q=1}^{3} m_{q}^{eff}$. Our prediction and comparison with other approaches are given in Table \ref{mag}.

\begin{table*}[ht]
{\begingroup
\caption{Magnetic moments {(in $\mu_N$)}} \label{mag}
\setlength{\tabcolsep}{9pt}
\renewcommand{\arraystretch}{1.5}
\begin{tabular}{c c c c c c}
\hline
\hline
State & Our & \cite{Ozdem:2024brk} & \cite{Patel:2007gx} & \cite{Gandhi:2019bju} & \cite{Kakadiya:2022zvy}  \\
\hline
$\Xi^{0}_c \frac{1}{2}^+$ & $-0.964$ & $0.41\pm0.05$ & $-0.932$ &$-1.011$ & $...$  \\
$\Xi^{0}_c \frac{3}{2}^+$ & $-0.748$ & $-0.51\pm0.05$  & $-0.671$& $-0.825$ & $...$   \\
$\Xi^{-}_b \frac{1}{2}^+$ & $-0.539$ & $-0.25\pm0.03$  & $-0.887$ & $...$ & $-0.963$  \\
$\Xi^{-}_b \frac{3}{2}^+$ & $-0.906$ & $-0.57\pm0.05$  & $-1.048$ & $...$ & $-1.499$ \\
\hline
\hline
\end{tabular}
\endgroup}
\end{table*}

\subsection{Radiative decay}\label{sec:3.1}

The electromagnetic radiative decay width can be expressed in terms of the radiative transition magnetic moment (in $ \mu_N $) and the photon energy ($ q = M_{3/2} - M_{1/2} $) as follows \cite{Dey:1994qi, Thakkar:2010ij}:  
\begin{eqnarray}
\Gamma_R = \frac{q^3}{4\pi} \frac{2}{2J+1} \frac{e^2}{m_p^2} |\mu_{\frac{3}{2}^+ \rightarrow \frac{1}{2}^+}|^2,
\end{eqnarray}  
where the transition magnetic moment is given by:  
\begin{multline}
\mu_{\frac{3}{2}^+ \rightarrow \frac{1}{2}^+} = \sum_i \big< \phi_{sf}^{\frac{3}{2}^+} | \mu_i \cdot \vec{\sigma_i} | \phi_{sf}^{\frac{1}{2}^+} \big> \\  
= \frac{\sqrt{2}}{\sqrt{3}} (\mu_d - \mu_s).
\end{multline}  

A key aspect of this calculation is the determination of the quark magnetic moments, which are obtained by taking the geometric mean of their effective masses. This is expressed as \cite{Thakkar:2010ij, Dhir:2009ax}:  
\begin{eqnarray}
m_i^{eff} = \sqrt{m_{i(\frac{3}{2}^+)}^{eff} \, m_{i(\frac{1}{2}^+)}^{eff}}.
\end{eqnarray}  

Our calculated Radiative decay width along with the transition magnetic moment is given in the table \ref{radiativedecay} 
\begin{table}[ht]
{\begingroup
\centering
\caption{Transition magnetic moment(TMM)(in $\mu_N$) and Decay Width(DW){(in keV)}} \label{radiativedecay}
\setlength{\tabcolsep}{5pt}
\renewcommand{\arraystretch}{1.5}
\begin{tabular}{ c c c c c  }
\hline
\hline
Baryon & TMM & DW & \cite{Kim:2021xpp} & \cite{Majethiya:2009vx}   \\
\hline
\textit{$\Xi^{*0}_c\rightarrow\Xi^0_c$} & $-0.056$ & $0.073$ & $0.080$& $0.300$ \\
\textit{$\Xi^{*-}_b\rightarrow\Xi^{-}_b$} & $-0.013$ & $0.003$ & $...$ &$0.090$ \\
\hline
\hline
\end{tabular}
\endgroup}
\end{table}
\subsection{Two body Weak decays of $\Xi_c^0$}
We adopt the methodology of Ref.\cite{Zhong:2022exp} to calculate the two body weak decays of $\Xi_c^0$.
A state-of-the-art effective Hamiltonian framework is employed to describe the weak decays of charmed baryons. At the quark level, the effective Hamiltonian governing the quark decay process $c \to \bar{q}_1 q_2 u$ is expressed as  
\begin{equation}
\begin{split}
\mathcal{H}_{\text{eff}} &= \frac{G_F}{\sqrt{2}} V_{q_1 c}^* V_{u q_2} 
\left(c_+ \mathcal{O}_+ + c_- \mathcal{O}_-\right) + h.c., \\
\mathcal{O}_+ &= \frac{1}{2}\left(\mathcal{O}_1 + \mathcal{O}_2\right), \qquad
\mathcal{O}_- = \frac{1}{2}\left(\mathcal{O}_1 - \mathcal{O}_2\right),
\end{split}
\end{equation}
where 
$\mathcal{O}_1 = (\bar{q}_1 c)(\bar{u} q_2)$ 
and 
$\mathcal{O}_2 = (\bar{u} c)(\bar{q}_1 q_2)$.  
The bilinear current is defined as  
\[
(\bar{q}_1 q_2) = \bar{q}_1 \gamma^\mu (1 - \gamma_5) q_2.
\]
The quark flavors appearing in the four-quark operator are denoted as  
$(q_1, q_2) = (s, d), \{(s, s), (d, d)\}, (d, s)$,  
corresponding respectively to Cabibbo-favored (CF), singly Cabibbo-suppressed (SCS), and doubly Cabibbo-suppressed (DCS) processes.

From the perspective of SU(3) flavor symmetry, the four-quark operators $\mathcal{O}_{\pm}$ in the weak Hamiltonian transform into two irreducible representations, $(\bm{6},\,\overline{\bm{15}})$ \cite{Savage:1989qr, Geng:2019xbo, Huang:2021aqu}, expressed as  
\begin{widetext}
\begin{equation}
\begin{split}
H(\overline{15})_k^{ij} &= 
\left(
\left(\begin{array}{ccc}
0&0&0\\0&0&0\\0&0&0
\end{array}\right),
\left(\begin{array}{ccc}
0&s_c&1\\s_c&0&0\\1&0&0
\end{array}\right),
\left(\begin{array}{ccc}
0&-s_c^2&-s_c\\-s_c^2&0&0\\-s_c&0&0
\end{array}\right)
\right),\\
H(6)_{ij} &= 
\left(\begin{array}{ccc}
0&0&0\\0&2&-2s_c\\0&-2s_c&2s_c^2
\end{array}\right),
\end{split}
\end{equation}
\end{widetext}
where $s_c=\sin\theta_c=\sqrt{0.23}$ \cite{ParticleDataGroup:2024cfk}, and $(i,j,k)=1,2,3$. Here, $H(\overline{15})$ is symmetric in the superscript indices, while $H(6)$ is symmetric in the subscript indices.
For two-body decays of antitriplet charmed baryons $B_c$ into octet baryons $B_n$ and mesons $M$, the amplitude is  
\[
\mathcal{M} = \langle M B_n | \mathcal{H}_{\text{eff}} | B_c \rangle 
= i\bar{u}_f(A - B\gamma_5)u_i,
\]
where $A$ and $B$ denote the $S$- and $P$-state amplitudes. These can be parameterized within SU(3) symmetry using irreducible tensor representations. The baryon, meson, and antitriplet matrices are  
\begin{widetext}
\[
\bm{B}_c = (\Xi_c^0, -\Xi_c^+, \Lambda_c^+),\quad
(\bB'_c)^{ij}=\epsilon^{ijk}(\bB_c)_k,
\]
\[
\bm{B}_n = 
\left(\begin{array}{ccc}
\frac{1}{\sqrt{6}}\Lambda^0+\frac{1}{\sqrt{2}}\Sigma^0 & \Sigma^+ & p\\
\Sigma^- & \frac{1}{\sqrt{6}}\Lambda^0-\frac{1}{\sqrt{2}}\Sigma^0 & n\\
\Xi^- & \Xi^0 & -\sqrt{\frac{2}{3}}\Lambda^0
\end{array}\right),
\]
\[
M =
\left(\begin{array}{ccc}
\frac{1}{\sqrt{2}}(\pi^0 + c_\phi\eta + s_\phi\eta') & \pi^+ & K^+\\
\pi^- & \frac{1}{\sqrt{2}}(-\pi^0 + c_\phi\eta + s_\phi\eta') & K^0\\
K^- & \bar{K}^0 & -s_\phi\eta + c_\phi\eta'
\end{array}\right),
\]
\end{widetext}
with $(c_\phi, s_\phi) = (\cos\phi, \sin\phi)$ and $\phi = 39.3^\circ$ \cite{Feldmann:1998vh, Feldmann:1998sh}.
To include SU(3) breaking effects, the amplitudes are extended through a dominant $\overline{3}$ representation \cite{Geng:2018bow},  
\begin{multline}
A' = u_1 (\bB_c)_i H(\overline{3})^i (\bB_n)^j_k(M)^k_j 
\\+ u_2 (\bB_c)_i H(\overline{3})^j (\bB_n)^i_k(M)^k_j 
\\+ u_3 (\bB_c)_i H(\overline{3})^j (\bB_n)^k_j(M)^i_k, \\
B' = A'|_{u_i \to v_i},
\end{multline}
with $H(\overline{3}) = (s_c, 0, 0)$.  
Combining total $S$- and $P$-wave amplitudes, $A = A_0 + A'$ and $B = B_0 + B'$, the decay width $\Gamma$ is 
\begin{multline}
\Gamma = \frac{p_c}{8\pi}
\Bigg[\frac{((m_i+m_f)^2 - m_P^2)}{m_i^2}|A|^2
\\+ \frac{((m_i-m_f)^2 - m_P^2)}{m_i^2}|B|^2\Bigg],
\end{multline}
where $m_i, m_f, m_P$ are the masses of the initial baryon, final baryon, and meson, respectively, and $p_c$ denotes the c.m. momentum in the baryon rest frame. Our calculated branching ratios, along with those from \cite{Zhong:2022exp} and experimental observations, are presented in the Table.\ref{tab:prediction}.

\begin{table*}
\caption{ The fitting results of amplitude coefficients in the irreducible amplitude approach
in the unit of $10^{-2} G_F {\rm GeV}^2$ obtained in ref.\cite{Zhong:2022exp}.} 
\label{tab:coeff}
\begin{center}
\setlength{\tabcolsep}{9pt}
\renewcommand\arraystretch{1.5}
{
\begin{tabular}
{ c c c c }
\hline
\hline
 Coefficient & Value &
 Coefficient & Value \\
\hline

$a_0$& $-1.20$&
$b_0$& $-0.70$\\
\hline

$a_1$& $-3.50$&
$b_1$& $8.60$\\
\hline

$a_2$& $1.45$&
$b_2$& $4.00$\\
\hline

$a_3$& $-1.98$&
$b_3$& $-0.80$\\
\hline

$a'_0$& $0.10$&
$b'_0$& $2.00$\\
\hline

$a_4$& $0.23$&
$b_4$& $-3.40$\\
\hline

$a_5$& $2.14$&
$b_5$& $3.00$\\
\hline

$a_6$& $1.56$&
$b_6$& $10.10$\\
\hline

$a_7$& $-0.59$&
$b_7$& $-0.90$\\
\hline

$u_1$& $0$&
$v_1$& $13.00$\\
\hline

$u_2$& $1.20$&
$v_2$& $-3.00$\\
\hline

$u_3$& $2.80$&
$v_3$& $4.80$\\
\hline
\end{tabular}
}
\end{center}
\end{table*}

\begin{table*}
\caption{The fitting results of all the weak decays of $\Xi_c^0$ into pseudoscalar mesons in SU(3) flavor symmetry breaking scenarios, including Cabibbo-favored, singly Cabibbo-suppressed and doubly Cabibbo-suppressed processes.}
\label{tab:prediction}
\begin{center}
\renewcommand\arraystretch{1.5}
\setlength{\tabcolsep}{3pt}
{
\begin{tabular}
{ l  c c c c c c}
\hline
\hline
Channel(CF)& $A/G_F$ &
{$A(10^{-1}G_F)$}&{$B(10^{-1}G_F)$} &Our($10^{2}\mathcal{B}$)&$10^{2}\mathcal{B}$\cite{Zhong:2022exp}&($10^{2}\mathcal{B}$)Exp.\cite{ParticleDataGroup:2024cfk}\\
\hline

$\Xi_c^0\to \Lambda^0 \overline{K}^0$ & $\frac{\sqrt{6}}{6}(-4a_1+2a_2+2a_3-2a_5+a_6+a_7)$ &
$0.390$&$-1.010$& $0.659$ &
$0.668$&-\\

$\Xi_c^0\to \Sigma^0 \overline{K}^0$ & $\frac{\sqrt{2}}{2}(-2a_2-2a_3+a_6-a_7)$&
$0.230$&$0.330$& $0.150$ &
$0.147$&-\\

$\Xi_c^0\to \Sigma^+ K^-$ & $2a_2+a_4+a_7$ &
$0.250$&$0.370$& $0.180$ &
$0.186$&$0.180\pm0.04$\\

$\Xi_c^0\to \Xi^0 \pi^0$ & $\frac{\sqrt{2}}{2}(-2a_1+2a_3+a_4-a_5)$&
$0.080$&$-1.780$& $0.773$&
$0.774$&$0.690\pm0.140$\\

$\Xi_c^0\to \Xi^0 \eta$ & \makecell[c]{$\frac{\sqrt{2}}{6}c_{\phi}(12a_0+6a_1-6a_3+6a'_0+3a_4+3a_5)$\\$+\frac{1}{3}s_{\phi}(-6a_0-6a_2-3a'_0-3a_7)$}&
$-0.290$&$0.580$& $0.242$&
$0.243$&$0.160\pm0.040$\\

$\Xi_c^0\to \Xi^0 \eta'$ & \makecell[c]{$\frac{\sqrt{2}}{6}s_{\phi}(12a_0+6a_1-6a_3+6a'_0+3a_4+3a_5)$\\$-\frac{1}{3}c_{\phi}(-6a_0-6a_2-3a'_0-3a_7)$}&
$-0.240$&$1.470$& $0.168$&
$0.163$&$0.110\pm0.040$\\

$\Xi_c^0\to \Xi^- \pi^+$ & $2a_1+a_5+a_6$&
$-0.330$&$3.030$& $2.434$&
$2.430$&$1.430\pm0.270$\\
\hline

Channel(SCS)& $s_c^{-1}$$A/G_F$&
{$A(10^{-2}G_F)$}&{$B(10^{-2}G_F)$}&Our($10^{4}\mathcal{B}$)&$10^{4}\mathcal{B}$\cite{Zhong:2022exp}&($10^{4}\mathcal{B}$)Exp.\cite{ParticleDataGroup:2024cfk}\\
\hline

$\Xi_c^0\to \Lambda^0 \pi^0$& \makecell[c]{$\frac{\sqrt{3}}{6}(-2a_1-2a_2+4a_3+3a_4-a_5-a_6-a_7$\\$+u_2+u_3)$}&
$-0.150$&$-3.180$& $3.758$&
$3.750$&-\\

$\Xi_c^0\to \Lambda^0 \eta$& \makecell[c]{$\frac{\sqrt{3}}{6}c_{\phi}(12a_0+2a_1+2a_2-4a_3+6a'_0+3a_4+a_5$\\$+a_6+a_7+2u_1+u_2+u_3)$\\$+\frac{\sqrt{6}}{6}s_{\phi}(-6a_0-4a_1-4a_2+2a_3-3a'_0-2a_5$\\$+a_6-2a_7+2u_1)$}&
$0.450$&$1.830$& $1.406$&
$1.400$&-\\

$\Xi_c^0\to \Lambda^0 \eta'$& \makecell[c]{$\frac{\sqrt{3}}{6}s_{\phi}(12a_0+2a_1+2a_2-4a_3+6a'_0+3a_4+a_5$\\$+a_6+a_7+2u_1+u_2+u_3)$\\$-\frac{\sqrt{6}}{6}c_{\phi}(-6a_0-4a_1-4a_2+2a_3-3a'_0-2a_5$\\$+a_6-2a_7+2u_1)$}&
$-0.770$&$4.090$& $3.119$&
$3.100$&-\\

$\Xi_c^0\to p K^-$& $-2a_2-a_4-a_7+u_1+u_3$&
$0.058$&$3.17$& $4.318$&
$4.31$&-\\

$\Xi_c^0\to n \overline{K}^0$& $2a_1-2a_2-2a_3+a_5-a_7+u_1$&
$-0.720$&$6.220$& $17.702$&
$17.730$&-\\

$\Xi_c^0\to \Sigma^0\pi^0$& \makecell[c]{$\frac{1}{2}(2a_1+2a_2-a_4+a_5-a_6+a_7$\\$+2u_1+u_2+u_3)$}&
$-0.038$&$5.430$& $9.239$&
$9.260$&-\\

$\Xi_c^0\to \Sigma^0 \eta$& \makecell[c]{$\frac{1}{2}c_{\phi}(-4a_0-2a_1-2a_2-2a'_0-a_4-a_5$\\$+a_6-a_7+u_2+u_3)$\\$+\frac{\sqrt{2}}{2}s_{\phi}(2a_0-2a_3+a'_0+a_6)$}&
$1.410$&$0.090$& $4.531$&
$4.540$&-\\

$\Xi_c^0\to \Sigma^0 \eta'$& \makecell[c]{$\frac{1}{2}s_{\phi}(-4a_0-2a_1-2a_2-2a'_0-a_4-a_5$\\$+a_6-a_7+u_2+u_3)$\\$-\frac{\sqrt{2}}{2}c_{\phi}(2a_0-2a_3+a'_0+a_6)$}&
$0.490$&$-2.440$& $0.913$&
$0.910$&-\\

$\Xi_c^0\to \Sigma^+ \pi^-$& $2a_2+a_4+a_7+u_1+u_3$&
$1.200$&$4.830$& $11.070$&
$11.080$&-\\

$\Xi_c^0\to \Sigma^- \pi^+$& $2a_1+a_5+a_6+u_1+u_2$&
$-0.470$&$9.050$& $25.943$&
$25.950$&-\\

$\Xi_c^0\to \Xi^0 K^0$& $-2a_1+2a_2+2a_3-a_5+a_7+u_1$&
$0.720$&$-0.380$& $1.213$&
$1.220$&-\\

$\Xi_c^0\to \Xi^- K^+$& $-2a_1-a_5-a_6+u_1+u_2$&
$1.010$&$-4.560$& $5.990$&
$5.990$&$3.900\pm1.100$\\
\hline

Channel(DCS)& $s_c^{-2}$$A/G_F$&
{$A(10^{-3}G_F)$}&{$B(10^{-3}G_F)$}&Our($10^{5}\mathcal{B}$)&$10^{5}\mathcal{B}$\cite{Zhong:2022exp}&($10^{5}\mathcal{B}$)Exp.\cite{ParticleDataGroup:2024cfk}\\
\hline

$\Xi_c^0\to \Lambda^0 K^0$& $\frac{\sqrt{6}}{6}(-2a_1+4a_2+4a_3-a_5-a_6+2a_7)$&
$0$&$-3.970$& $0.467$&
$0.468$&-\\

$\Xi_c^0\to p \pi^-$& $-2a_2-a_4-a_7$&
$-1.280$&$-1.870$& $0.585$&
$0.588$&-\\

$\Xi_c^0\to n \pi^0$& $\frac{\sqrt{2}}{2}(2a_2-a_4+a_7)$&
$0.740$&$3.740$& $0.840$&
$0.845$&-\\

$\Xi_c^0\to n \eta$& \makecell[c]{$\frac{\sqrt{2}}{2}c_{\phi}(-4a_0-2a_2-2a'_0-a_4-a_7)$\\$+s_{\phi}(2a_0+2a_1-2a_3+a'_0+a_5)$}&
$-0.450$&$5.800$& $1.424$&
$1.430$&-\\

$\Xi_c^0\to n \eta'$&\makecell[c]{$\frac{\sqrt{2}}{2}s_{\phi}(-4a_0-2a_2-2a'_0-a_4-a_7)$\\$-c_{\phi}(2a_0+2a_1-2a_3+a'_0+a_5)$}&
$1.710$&$-9.850$& $2.774$&
$2.77$&-\\

$\Xi_c^0\to \Sigma^0 K^0$&$\frac{\sqrt{2}}{2}(2a_1+a_5-a_6)$&
$-2.290$&$3.600$& $1.545$&
$1.550$&-\\

$\Xi_c^0\to \Sigma^- K^+$&$-2a_1-a_5-a_6$&
$1.660$&$-15.280$& $6.375$&
$6.370$&-\\
\hline
\end{tabular}
}
\end{center}
\end{table*}

\subsection{Non leptonic decays of $\Xi^-_b$ baryons}
The color-allowed two-body nonleptonic decays of bottom baryons $\Xi^-_b$ with the emission of a pseudoscalar ($\pi^{-}$, $K^{-}$, $D^{-}$, and $D_s^{-}$) or a vector meson ($\rho^{-}$, $K^{*-}$, $D^{*-}$, and $D_s^{*-}$) are calculated within na\"{i}ve factorization approach. Under this approximation, the hadronic transition matrix element is factorized into a product of two independent matrix elements \cite{Ke:2019smy}. Accordingly, we can express, 
\begin{multline}
\langle\mathcal{B}_c^{(*)}(P^{\prime},J_{z}^{\prime}) \ M^{-} \ |\mathcal{H}_{\text{eff}}| \ \mathcal{B}_b(P,J_{z})\rangle\\
=\frac{G_F}{\sqrt{2}} V_{cb} V_{qq^{\prime}}^{*}
\langle M^{-}|\bar{q}^{\prime}\gamma_{\mu}(1-\gamma_{5})q|0\rangle\\
\times\langle\mathcal{B}_c^{(*)}(P^{\prime},J_{z}^{\prime})|\bar{c}\gamma^{\mu}(1-\gamma_{5})b|\mathcal{B}_b(P,J_{z})\rangle,
\end{multline}
where the meson transition term is given by
\begin{equation}
\langle M|\bar{q}^{\prime}\gamma_{\mu}(1-\gamma_{5})q|0\rangle
=\left\{
\begin{array}{ll}
if_{P}q_{\mu}, &M=P\\
if_{V}\epsilon_{\mu}^{*}m_{V}, &M=V
\end{array}.
\right.
\end{equation}
In the case where the final state includes a pseudoscalar meson, the decay width \cite{Ke:2019smy} takes the following form 
\begin{multline}
\Gamma=\frac{|p_c|}{8\pi}\Big(\frac{(M+M^{\prime})^2-m^2}{M^2}|A|^2\\+\frac{(M-M^{\prime})^2-m^2}{M^2}|B|^2\Big),
\label{eq:PhysicsP}
\end{multline}
\begin{equation}
    \alpha=\frac{2\kappa\text{Re}(A^*B)}{|A|^2+\kappa^2|B|^2},
\end{equation}
whereas for the transitions emitting vector meson in final states \cite{Ke:2019smy} we can write 
\begin{multline}
\Gamma=\frac{|p_c|(E^{\prime}+M^{\prime})}{4\pi M}\Big(2(|S|^2+|P_2|^2)\\+\frac{E_m^2}{m^2}(|S+D|^2+|P_1|^2)\Big),
\label{eq:PhysicsV} 
\end{multline}
\begin{eqnarray}
    \alpha=\frac{4m^2\text{Re}(S^*P_2)+2E_m^2\text{Re}(S+D)^*P_1}{2m^2(|S|^2+|P_2|^2)+E_m^2(|S+D|^2+|P_1|^2)},
\end{eqnarray}
Here, $p_c$ represents the momentum of the daughter baryon measured in the rest frame of the parent baryon, and $\kappa = |p_c|/(E^{\prime}+M^{\prime})$. Additionally, $M$ ($E$) and $M'$ ($E'$) denote the masses (energies) of the parent and daughter baryons, respectively, while $m$ ($E_m$) corresponds to the mass (energy) of the final-state meson. The amplitudes $A$ and $B$ in Eqs. (\ref{eq:PhysicsP}) are given by
\begin{equation}
A=\frac{G_F}{\sqrt{2}}V_{cb}V_{qq^{\prime}}^*a_{i}f_{P}(M-M^{\prime})f_1^V(m^2),
\end{equation}
\begin{equation}
B=-\frac{G_F}{\sqrt{2}}V_{cb}V_{qq^{\prime}}^*a_{i}f_{P}(M+M^{\prime})g_1^A(m^2),
\end{equation}
and $S$, $P_{1,2}$ and $D$ in Eqs. (\ref{eq:PhysicsV}) are expressed as
\begin{equation}
S=A_1,    
\end{equation}
\begin{equation}
P_1=-\frac{|p_c|}{E_m}\left(\frac{M+M^{\prime}}{E^{\prime}+M^{\prime}}B_1+MB_2\right),    
\end{equation}
\begin{equation}
P_2=\frac{|p_c|}{E^{\prime}+M^{\prime}}B_1,    
\end{equation}
\begin{eqnarray}
D=\frac{|p_c|^2}{E_m(E^{\prime}+M^{\prime})}(A_1-MA_2)    
\end{eqnarray}
with
\begin{multline}
    A_{1}=\frac{G_F}{\sqrt{2}}V_{cb}V_{qq^{\prime}}^*a_{i}f_{V}m_{V}\Big(g_1^A(m^2)\\+g_2^A(m^2)\frac{M-M^{\prime}}{M}\Big),
\end{multline}
\begin{eqnarray}
    A_{2}=\frac{G_F}{\sqrt{2}}V_{cb}V_{qq^{\prime}}^*a_{i}f_{V}m_{V}\left(2g_{2}^A(m^2)\right),
\end{eqnarray}
\begin{multline}
B_{1}=\frac{G_F}{\sqrt{2}}V_{cb}V_{qq^{\prime}}^*a_{i}f_{V}m_{V}\Big(f_1^V(m^2)\\-f_2^V(m^2)\frac{M+M^{\prime}}{M}\Big)    
\end{multline}
\begin{equation}
B_{2}=\frac{G_F}{\sqrt{2}}V_{cb}V_{qq^{\prime}}^*a_{i}f_{V}m_{V}\left(2f_{2}^V(m^2)\right),
\end{equation}
For numerical evaluation, the values of the form factors we adopted from the three-body light front quark model from Ref. \cite{Li:2021kfb}. For the color allowed transition, we consider the coefficient $a_1 = 1.018$  \cite{Chua:2019yqh}. The lifetime is taken as 
$\tau_{\Xi_b^-} = 1.570~\text{ps}$ and $\tau_{\Xi_b^0} = 1.477~\text{ps}$\cite{ParticleDataGroup:2024cfk}. The CKM matrix elements used as \cite{ParticleDataGroup:2024cfk} 
\begin{equation*}
\begin{split}
& V_{cb}=0.0405,\ V_{ud}=0.9740,\ V_{us}=0.2265,\\
& V_{cd}=0.2264,\ V_{cs}=0.9732.
\end{split}
\end{equation*}
The decay constants of pseudoscalar and vector mesons are taken from \cite{ParticleDataGroup:2024cfk}
\begin{equation*}
\begin{split}
&f_{\pi}=130.2,\ f_{K}=155.6,\ f_{D}=211.9,\ f_{D_s}=249.0,\\
&f_{\rho}=216,\ f_{K^*}=210,\ f_{D^*}=220,\ f_{D_s^*}=230.
\end{split}
\end{equation*}

In the present work, the branching ratios are predicted using the baryon masses calculated within our formalism. Table \ref{tab:Comparison} compares our results for the branching fractions of $\Xi_b^{-} \to \Xi_c^{0}M^-$ decays with those from various theoretical models, including the nonrelativistic quark model \cite{Cheng:1996cs}, the relativistic three-quark model \cite{Ivanov:1997hi,Ivanov:1997ra}, the light-front quark model \cite{Zhao:2018zcb,Chua:2019yqh}, and the covariant confined quark model \cite{Gutsche:2018utw}. This comparison highlights both the consistency and differences between our predictions and those of previous studies.

\begin{table*}[htbp]\centering
\caption{The branching ratio for $(\Xi_b^{-}\to\Xi_c^{0}M^-)$. All the values are multiplied by a factor of $10^{-3}$.}
\label{tab:Comparison}
\renewcommand\arraystretch{1.5}
\setlength{\tabcolsep}{4.5pt}
\begin{tabular}{l c c c c c c c}
\hline
\hline
& Present & \cite{Li:2021kfb}      & \cite{Cheng:1996cs}    & \cite{Ivanov:1997hi,Ivanov:1997ra}    & \cite{Zhao:2018zcb}  &  \cite{Gutsche:2018utw}  & \cite{Chua:2019yqh}  \\
$\Xi_b^{-}\to\Xi_c^{0}\pi^-$  &  4.02\ (4.27) & 4.03\ (4.29) &4.9\ (5.2)  &7.08\ (10.13)   &8.37\ (8.93)   &$-$   &$3.66^{+2.29}_{-1.59}$\ ($3.88^{+2.43}_{-1.69}$)     \\

$\Xi_b^{-}\to\Xi_c^{0}K^-$   & 0.30\ (0.32)  & 0.31\ (0.33) &$-$         &$-$             &0.667\ (0.711) &$-$   &$0.28^{+0.17}_{-0.12}$\ ($0.29^{+0.18}_{-0.13}$)     \\

$\Xi_b^{-}\to\Xi_c^{0}D^-$   &  0.46\ (0.53) &0.58\ (0.62) &$-$         &$-$             &0.949\ (1.03)  &0.45  &$0.43^{+0.29}_{-0.20}$\ ($0.45^{+0.31}_{-0.21}$)     \\
$\Xi_b^{-}\to\Xi_c^{0}D_s^-$  & 10.5\ (11.2 ) & 14.8\ (15.7) &14.6        &$-$             &24.6\ (26.2)   &$-$   &$10.87^{+7.51}_{-5.03}$\ ($11.54^{+7.98}_{-5.34}$)   \\
\hline
$\Xi_b^{-}\to\Xi_{c}^{0}(2S) \pi^-$    & 0.98 \ (0.92) &  1.78\  (1.89 ) & \\

$\Xi_b^{-}\to\Xi_c^{0}(2S)K^-$  & 0.04\ (0.05)   &  0.04\ (0.05)  \\

$\Xi_b^{-}\to\Xi_c^{0}(2S)D^-$   & 0.05\ (0.05)  &   0.04\ (0.05)  \\
$\Xi_b^{-}\to\Xi_c^{0}(2S)D_s^-$   & 1.01\ (1.03) & 1.05\ (1.12) \\
\hline
 $\Xi_b^{-}\to\Xi_c^{0}\rho^-$ & 11.2\ (12.5) &13.3\ (14.1) &$-$         &$-$             &24.0\ (25.6)   &$-$   &$10.88^{+6.83}_{-4.74}$\ ($11.56^{+7.25}_{-5.04}$)   \\
$\Xi_b^{-}\to\Xi_c^{0}K^{*-}$ & 0.49\ (0.56 ) &0.71\ (0.76) &$-$         &$-$             &1.23\ (1.31)   &$-$   &$0.56^{+0.35}_{-0.24}$\ ($0.60^{+0.37}_{-0.26}$)     \\
$\Xi_b^{-}\to\Xi_c^{0}D^{*-}$ & 1.03\ ( 1.07) &1.51\ (1.60) &$-$         &$-$             &1.54\ (1.64)   &0.95  &$0.77^{+0.50}_{-0.35}$\ ($0.82^{+0.53}_{-0.37}$)     \\

$\Xi_b^{-}\to\Xi_c^{0}D_s^{*-}$ & (30.1)\ (32.8) &32.4\ (34.4) &23.1        &$-$             &36.5\ (39.0)   &$-$   &$16.24^{+10.54}_{-7.25}$\ ($17.26^{+11.2}_{-7.70}$)  \\

\hline
$\Xi_b^{-}\to\Xi_c^{0}(2S)\rho^-$  & 2.37\ (2.86) & 2.78 \ (2.95)  \\
$\Xi_b^{-}\to\Xi_c^{0}(2S)K^{*-}$  & 0.08\ (0.09) &  0.09 \ (0.10) \\
$\Xi_b^{-}\to\Xi_c^{0}(2S)D^{*-}$  & 0.08\ (0.09) & 0.12 \ (0.12) \\
$\Xi_b^{-}\to\Xi_c^{0}(2S)D_s^{*-}$ & 2.14 \ (2.22) &  2.30 \ (2.45)\\
\hline
\hline
\end{tabular}
\end{table*}

\section{regge trajectories}\label{regge}
\begin{figure}[hbt!]
    \includegraphics[width=0.5\textwidth]{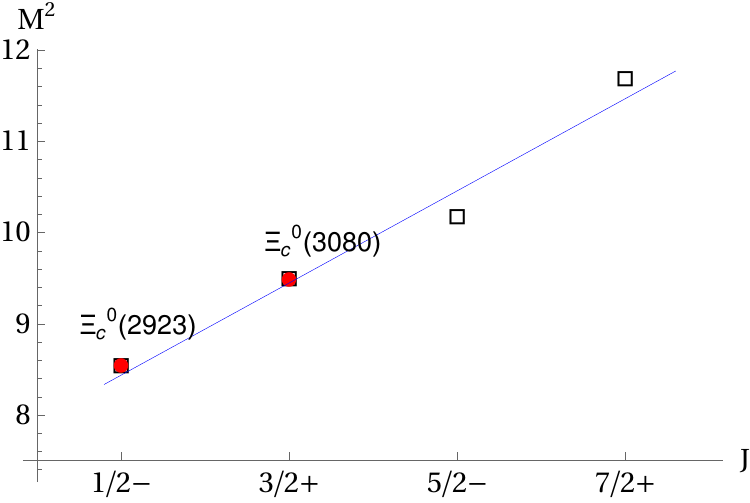}
    \caption{Regge trajectory for $2^2P_{\frac{1}{2}}$, $2^4D_{\frac{3}{2}}$, $2^2F_{\frac{5}{2}}$, \& $2^4D_{\frac{7}{2}}$ state masses. Red dot represents the experimental value of $\Xi_c^0(2923)$ and $\Xi_c^0(3080)$.}
    \label{image1}
\end{figure}

\begin{figure}[hbt!]
    \includegraphics[width=0.5\textwidth]{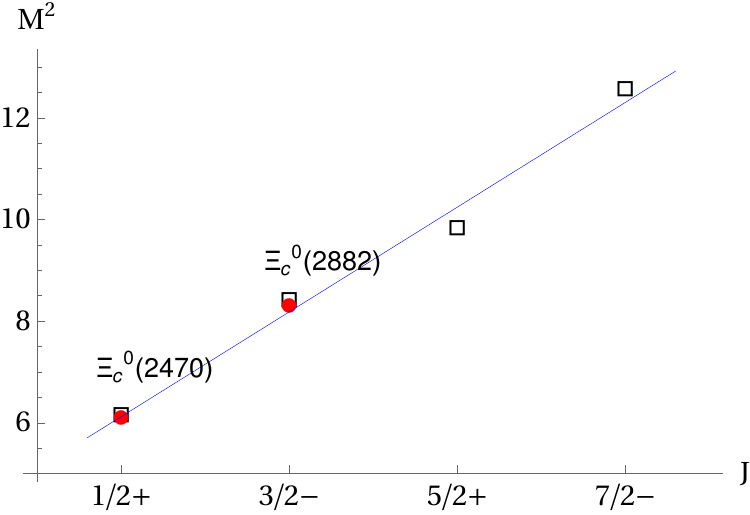}
    \caption{Regge trajectory for $1^2S_{\frac{1}{2}}$, $1^2P_{\frac{3}{2}}$, $1^4D_{\frac{5}{2}}$, \& $1^4F_{\frac{7}{2}}$ state masses. Red dots represent the experimental value of $\Xi_c^0(2470)$ and $\Xi_c^0(2882)$.}
    \label{image2}
\end{figure}

\begin{figure}[hbt!]
    \includegraphics[width=0.5\textwidth]{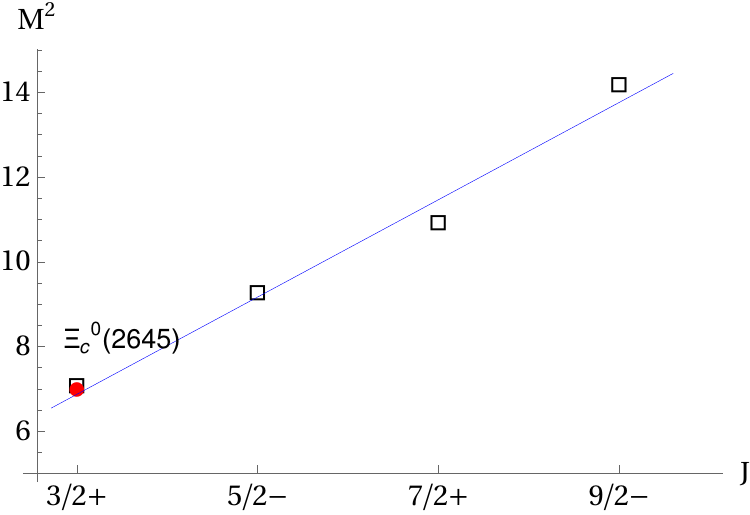}
    \caption{Regge trajectory for $1^4S_{\frac{3}{2}}$, $1^4P_{\frac{5}{2}}$, $1^4D_{\frac{7}{2}}$, \& $1^4F_{\frac{9}{2}}$ state masses.  Red dot represents the experimental values of $\Xi_c^0(2645)$.}
    \label{image3}
\end{figure}

\begin{figure}[hbt!]
    \includegraphics[width=0.5\textwidth]{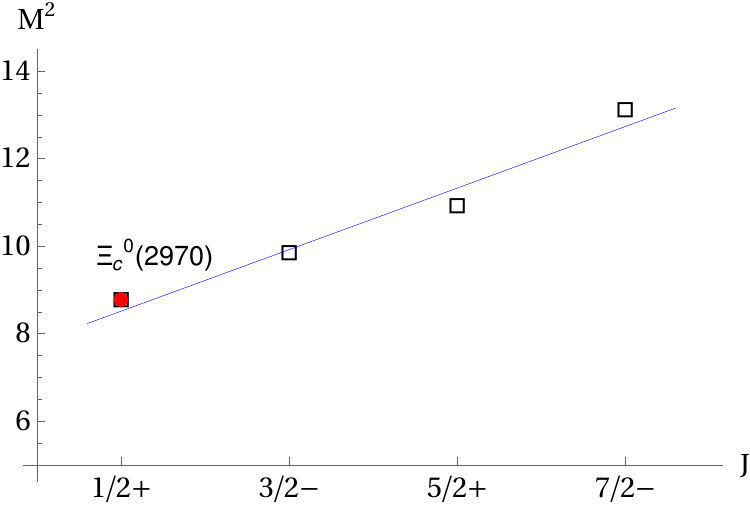}
    \caption{Regge trajectory for $2^2D_{\frac{1}{2}}$, $2^2P_{\frac{3}{2}}$, $2^4D_{\frac{5}{2}}$, \& $2^2F_{\frac{7}{2}}$ state masses. Red dot represents the experimental value of $\Xi_c^0(2970)$}.
    \label{image4}
\end{figure}

\begin{figure}[hbt!]
    \includegraphics[width=0.5\textwidth]{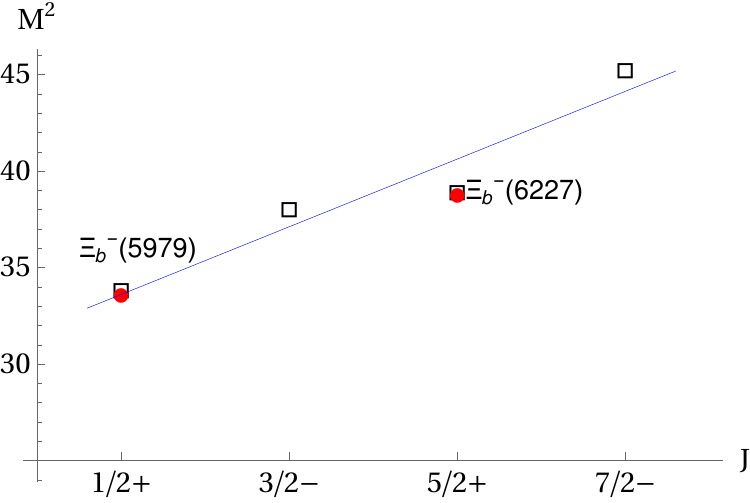}
    \caption{Regge trajectory for $1^2S_{\frac{1}{2}}$, $1^2P_{\frac{3}{2}}$, $1^4D_{\frac{5}{2}}$, \& $1^4F_{\frac{7}{2}}$ state masses. Red dots represent the experimental value of $\Xi_b^-(5979)$ and $\Xi_b^-(6227)$.}
    \label{image5}
\end{figure}

\begin{figure}[hbt!]
    \includegraphics[width=0.5\textwidth]{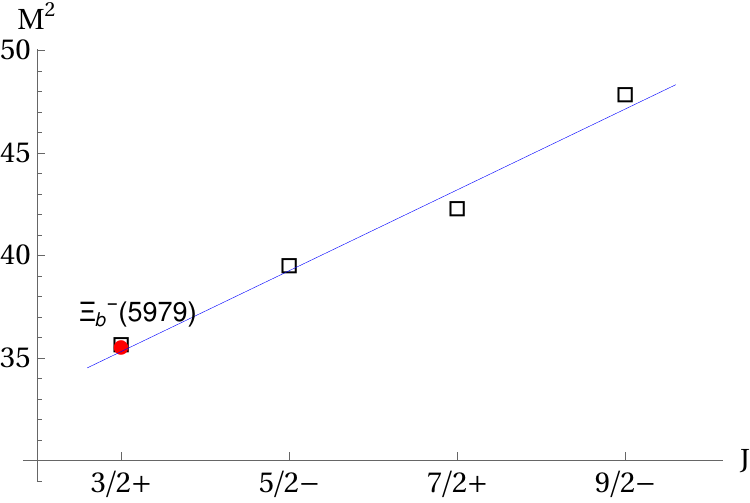}
    \caption{Regge trajectory for $1^4S_{\frac{3}{2}}$, $1^4P_{\frac{5}{2}}$, $1^4D_{\frac{7}{2}}$, \& $1^4F_{\frac{9}{2}}$ state masses.  Red dot represents the experimental values of $\Xi_b^-(5979)$.}
    \label{image6}
\end{figure}

Regge trajectories provide a valuable framework for assigning quantum numbers, such as spin, parity, and other intrinsic properties, to hadronic states. Having determined the masses of orbitally and radially excited singly heavy baryons up to high excitation levels, we construct the corresponding Regge trajectories in the $(J, M^{2})$ plane. These trajectories are obtained by fitting the linear relation
\begin{equation}
    J = \alpha M^{2} + \alpha_{0},
\end{equation}
where $\alpha$ and $\alpha_{0}$ denote the slope and intercept, respectively. The resulting plots are shown in Fig.~\ref{image1}, \ref{image2}, \ref{image3}, \ref{image4}, \ref{image5} and \ref{image6}, and the extracted parameters are summarized in Table~\ref{tab:8}.

\begin{table}[!tbh]
{
\begingroup
\caption{Slope($\alpha$) and intercept($\alpha_0$)} \label{tab:8}
\setlength{\tabcolsep}{9pt}
\renewcommand{\arraystretch}{1.5}
\begin{tabular}{  c c c   }
\hline
\hline
Baryon State &Slope & Intercept\\
\hline
$\Xi_c^{0} \frac{1}{2}^-$  & $0.990\pm0.041$ & $-7.848\pm1.529$\\
$\Xi_c^{0} \frac{1}{2}^+$  & $0.485\pm0.020$ & $-2.468\pm0.481$\\
$\Xi_c^{0} \frac{3}{2}^+$ & $0.436\pm0.018$ & $-1.493\pm0.291$ \\
$\Xi_c^{0} \frac{1}{2}^+$  & $0.711\pm0.029$ & $-5.553\pm1.082$\\
$\Xi_b^{-} \frac{1}{2}^+$  & $0.285\pm0.020$ & $-9.077\pm0.395$\\
$\Xi_b^{-} \frac{3}{2}^+$  & $0.253\pm0.018$ & $-7.452\pm0.311$\\
\hline
\end{tabular}
\endgroup}
\end{table}

\section{Discussion and Conclusion}\label{sec:4}
Reducing the number of free parameters is essential for strengthening the predictive power of any particle-physics model. In this work, we carried out a parameter-reduction analysis of the Independent Quark Model (IQM) within the relativistic Dirac framework, utilizing extracted potential parameters for a wide range of baryons previously.

We have recently extended the IQM—originally developed for mesons—to several baryonic systems \cite{Patel:2023wbs, Patel:2025gbw, Patel:2024wfo, Shah:2023myt, Shah:2023mzg}. Our earlier studies successfully described the spectroscopy and decay properties of $\Xi^{0}$, $\Omega_c^{0}$, $\Omega_b^{-}$, $\Omega_{cc}^{+}$, and $\Xi_{cc}^{++}$ \cite{Patel:2025gbw, Patel:2024wfo, Shah:2023myt, Shah:2023mzg}. In particular, we reproduced the mass spectra, magnetic moments, and radiative decay modes of the $\Xi_c^{0}$ baryon \cite{Patel:2024wfo} and proposed spin–parity assignments for several of its excited states. Similarly, our study of $\Omega_c^{0}$ and $\Omega_b^{-}$ \cite{Patel:2025gbw} yielded consistent predictions for their radial and orbital excitations and provided assignments for the experimentally observed $\Omega_c^{0}(3000)$, $\Omega_c^{0}(3050)$, $\Omega_c^{0}(3067)$, $\Omega_c^{0}(3120)$, and $\Omega_c^{0}(3185)$ states.

Our combined analysis of singly and doubly heavy baryons revealed a clear linear correlation among the fitted potential parameters. As shown in Table~\ref{tab:table1} and Fig.~\ref{fig1}, the parameters follow the relation
\[
\Lambda = -0.7308\,V_0 + 0.5239,
\]
allowing us to express one parameter in terms of the other and thereby reduce the model to a single free potential parameter. We tested this reduction strategy on the intermediate-mass baryons $\Xi_c^{0}$ and $\Xi_b^{-}$, where precise experimental data are available, and found that the reduced-parameter model successfully reproduces their spectroscopic and decay properties. This confirms the robustness and predictive efficiency of the parameter-reduced Independent Quark Model.

Our predicted masses, along with the available experimental data and other theoretical predictions, are presented in Tables~\ref{sc}, \ref{pc}, \ref{dc}, and \ref{fc} for the $S$, $P$, $D$, and $F$ states of the $\Xi_c^0$ baryon, and in Tables~\ref{sb}, \ref{pb}, \ref{db}, and \ref{fb} for the corresponding states of the $\Xi_b^-$ baryon. Our predicted ground state and first radial excited state masses are in good agreement with observed data\cite{ParticleDataGroup:2024cfk} and lattice QCD predictions as well
\cite{Brown:2014ena}. We predict the masses for orbital excited states of $\Xi_c^0$ and $\Xi_b^-$ to be of the order of a hundred $MeV$ less than the theoretical predictions \cite{Shah:2016mig, Gandhi:2019bju, Ebert:2011kk, Oudichhya:2021yln}.

Using the Regge trajectories, we predict the spin parity for some of the observed states. We predict that the spin parity for the observed states $\Xi_c^0(2923)$, $\Xi_c^0(3080)$, $\Xi_c^0(2882)$, and $\Xi_c^0(2970)$ could be $\frac{1}{2}^-$, $\frac{3}{2}^+$, $\frac{3}{2}^-$, and $\frac{1}{2}^+$. Those correspond to $2^2P_{\frac{1}{2}}$, $2^4D_{\frac{3}{2}}$, $1^2P_{\frac{3}{2}}$, and $2^2D_{\frac{1}{2}}$. Our predictions for spin parity agree with observational evidence, especially for $\Xi_c^0(2970)$\cite{Belle:2020tom}. For the $\Xi_b^-$ baryon, we predict that the spin parity for the observed state $\Xi_b^-(6227)$ could be $\frac{5}{2}^+$, which corresponds to $1^4D_{\frac{5}{2}}$. We also infer that the spin parity for the observed state $\Xi_b^-(6100)$ could be $\frac{3}{2}^+$, which corresponds to $2^4F_{\frac{3}{2}}$; however, this requires further investigation. 

We have calculated the magnetic moments for both the ground state and the first radially excited states. Our results, along with those obtained from various other theoretical approaches, are presented in Table \ref{mag}. While our predicted magnetic moments exhibit noticeable deviations from earlier theoretical estimates, current experimental data are insufficient to confirm or refute these differences. Given the prior success of our model in predicting related observables, we anticipate that future measurements will align with our findings.
Similarly, the computed radiative decay widths are listed in Table \ref{radiativedecay}. Despite their relatively small values, these decay processes play an essential role in probing the electromagnetic structure of heavy baryons, which motivates our detailed calculation.

We used the methodology of Ref.~\cite{Zhong:2022exp} to evaluate the two-body weak decays of \(\Xi_c^0\). Within this approach, the decays are described through an effective Hamiltonian framework in which the quark-level transition \(c \to \bar{q}_1 q_2 u\) is governed by four-quark operators classified as Cabibbo-favored, singly Cabibbo-suppressed, or doubly Cabibbo-suppressed processes.
The operators in the effective Hamiltonian are organized into the $SU(3)$ flavor representations \(\mathbf{6}\) and \(\overline{\mathbf{15}}\) \cite{Savage:1989qr, Geng:2019xbo, Huang:2021aqu}, and the baryons and mesons appearing in the final state are expressed through the standard $SU(3)$ matrices with the physical \(\eta-\eta'\) mixing parameters taken from Refs.~\cite{Feldmann:1998vh, Feldmann:1998sh}. The $SU(3)$-breaking effects are incorporated through the dominant \(\overline{\mathbf{3}}\) contribution following Ref.~\cite{Geng:2018bow}.
The total amplitudes are obtained by combining the $SU(3)$-symmetric and $SU(3)$-breaking parts, and the decay widths are then computed using the usual two-body formula. The resulting branching ratios, along with those from Ref.~\cite{Zhong:2022exp} and the available experimental data, are presented in Table~\ref{tab:prediction}.
Our predicted branching ratios for the decays 
$\Xi_c^0 \to \Sigma^+ K^-$, 
$\Xi_c^0 \to \Xi^0 \pi^0$, 
$\Xi_c^0 \to \Xi^0 \eta$, 
and $\Xi_c^0 \to \Xi^0 \eta'$ show very good agreement with the available experimental measurements.  
In contrast, the branching ratios obtained for 
$\Xi_c^0 \to \Xi^- \pi^+$ and $\Xi_c^0 \to \Xi^- K^+$ are somewhat higher than the observed values. 
This deviation is likely attributed to the fitted parameters used in Ref.~\cite{Zhong:2022exp}, which may be further refined with improved data in future studies.

For the two-body nonleptonic decays of \(\Xi^-_b\), we adopted the methodology of Ref.~\cite{Ke:2019smy}.
In this approach, the color-allowed transitions are treated within the naïve factorization framework, in which the hadronic matrix element is factorized into a product of a meson emission matrix element and a baryonic transition matrix element. The meson emission terms are expressed through the corresponding decay constants, whereas the baryonic transition is described using the vector and axial–vector form factors.
Under this formalism, the decay widths and angular asymmetries for final states containing either a pseudoscalar or a vector meson follow the expressions given in Ref.~\cite{Ke:2019smy}. The amplitudes involved are written in terms of the form factors, decay constants, and kinematic quantities characterizing the parent and daughter baryons and the emitted meson. For numerical evaluation, the transition form factors are taken from the three-body light-front quark model of Ref.~\cite{Li:2021kfb}. The baryon lifetimes, CKM matrix elements, and meson decay constants are adopted from Ref.~\cite{ParticleDataGroup:2024cfk}.
Using this methodology, the branching ratios of \(\Xi_b^{-} \to \Xi_c^{0} M^-\) are predicted with baryon masses obtained from our formalism. A comparison with the results from other theoretical approaches \cite{Cheng:1996cs, Ivanov:1997hi, Ivanov:1997ra, Zhao:2018zcb, Chua:2019yqh, Gutsche:2018utw} is presented in Table~\ref{tab:Comparison}. This comparison illustrates both the agreement and distinctions between the present predictions and earlier theoretical studies.

All the results indicate the effectiveness of our parameter-reduction procedure within the independent quark model.  
The resulting linear relation can now be employed to determine the model parameters for other baryons and to carry out their spectroscopy.  
This direction will be explored in future work.  

\begin{acknowledgments}
The financial support from the University Grants Commission (UGC-India) under the Savitribai Jyotirao Phule Single Girl Child Fellowship (SJSGC) scheme, Ref. No.(F. No. 82-7/2022(SA-III)) is gratefully acknowledge by Ms. Rameshri Patel. The authors sincerely acknowledge prof. P.C.Vinodkumar for his insightful knowledge imparted during this work.
\end{acknowledgments}


\end{document}